\documentclass{article}\usepackage[]{graphicx}\usepackage[]{color}
\makeatletter
\def\maxwidth{ %
  \ifdim\Gin@nat@width>\linewidth
    \linewidth
  \else
    \Gin@nat@width
  \fi
}
\makeatother

\definecolor{fgcolor}{rgb}{0.345, 0.345, 0.345}
\newcommand{\hlnum}[1]{\textcolor[rgb]{0.686,0.059,0.569}{#1}}%
\newcommand{\hlstr}[1]{\textcolor[rgb]{0.192,0.494,0.8}{#1}}%
\newcommand{\hlcom}[1]{\textcolor[rgb]{0.678,0.584,0.686}{\textit{#1}}}%
\newcommand{\hlopt}[1]{\textcolor[rgb]{0,0,0}{#1}}%
\newcommand{\hlstd}[1]{\textcolor[rgb]{0.345,0.345,0.345}{#1}}%
\newcommand{\hlkwa}[1]{\textcolor[rgb]{0.161,0.373,0.58}{\textbf{#1}}}%
\newcommand{\hlkwb}[1]{\textcolor[rgb]{0.69,0.353,0.396}{#1}}%
\newcommand{\hlkwc}[1]{\textcolor[rgb]{0.333,0.667,0.333}{#1}}%
\newcommand{\hlkwd}[1]{\textcolor[rgb]{0.737,0.353,0.396}{\textbf{#1}}}%

\usepackage{framed}
\makeatletter
\newenvironment{kframe}{%
 \def\at@end@of@kframe{}%
 \ifinner\ifhmode%
  \def\at@end@of@kframe{\end{minipage}}%
  \begin{minipage}{\columnwidth}%
 \fi\fi%
 \def\FrameCommand##1{\hskip\@totalleftmargin \hskip-\fboxsep
 \colorbox{shadecolor}{##1}\hskip-\fboxsep
     \hskip-\linewidth \hskip-\@totalleftmargin \hskip\columnwidth}%
 \MakeFramed {\advance\hsize-\width
   \@totalleftmargin\z@ \linewidth\hsize
   \@setminipage}}%
 {\par\unskip\endMakeFramed%
 \at@end@of@kframe}
\makeatother

\definecolor{shadecolor}{rgb}{.97, .97, .97}
\definecolor{messagecolor}{rgb}{0, 0, 0}
\definecolor{warningcolor}{rgb}{1, 0, 1}
\definecolor{errorcolor}{rgb}{1, 0, 0}
\newenvironment{knitrout}{}{} 

\usepackage{alltt}

\usepackage{rotating, graphicx, overpic, wrapfig, psfrag, amsfonts,
  amsmath, amssymb, amsthm, color, fullpage, url, natbib, caption,
  subcaption, authblk, color, hyperref, bm, blkarray, array, tikz, 
  tkz-euclide, verbatim}
\usepackage[boxed]{algorithm2e}

\newcommand{\al}{\alpha}

\newcommand{\f}[2]{\frac{#1}{#2}}

\newcommand{\mc}[1]{\mathcal{#1}}

\newcommand{\PRRV}[1]{P\left[#1\right]}

\newcommand{\set}[1]{\left\{#1\right\}}

\newcommand{\C}{\mathbb{C}}
\newcommand{\R}{\mathbb{R}}
\newcommand{\Q}{\mathbb{Q}}
\newcommand{\N}{\mathbb{N}}
\newcommand{\Z}{\mathbb{Z}}
\newcommand{\F}{\mathbb{F}}
\newcommand{\K}{\mathbb{K}}

\newcommand{\varty}[1]{\mc{V}\left(#1\right)}

\renewcommand{\hbar}{\mbox{------}}

\newcommand{\ma}[1]{{\bf{#1}}}
\newcommand{\ve}[1]{\textbf{#1}}
\newcommand{\veg}[1]{\bm{#1}}

\newcommand{\defn}[1]{\color{red}{\emph{#1}}\color{black}{}}

\definecolor{BUgreen}{RGB}{24,144,41} 
\definecolor{mmGreen}{RGB}{61,153,86} 
\definecolor{darkBlue}{RGB}{0,92,141}

\definecolor{niceBlue}{RGB}{13,94,141}
\definecolor{niceRed}{RGB}{177,6,16}

\definecolor{darkgreen}{RGB}{0,120,36}
\definecolor{darkblue}{RGB}{0,59,114}
\hypersetup{colorlinks,breaklinks,
            linkcolor=darkblue,urlcolor=darkblue,
            anchorcolor=darkblue,citecolor=darkblue}

\definecolor{exColor}{RGB}{0,92,141}

\definecolor{propColor}{RGB}{24,144,41}

\definecolor{pfColor}{RGB}{153,139,61}

\definecolor{noteColor}{RGB}{24,144,41} 

\newcommand{\code}[1]{\tt #1\fontfamily{\rmdefault}\selectfont}
\newcommand{\pl}[1]{{\fontfamily{\sfdefault}\selectfont #1}}

\newcommand{\pkg}[1]{\textbf{\pl{#1}}}

\definecolor{darkgreen2}{RGB}{61,153,86}
\definecolor{gray}{RGB}{240, 240, 240}

\tikzset{
  yn/.style={draw,thick,rounded corners,fill=yellow!20,inner sep=.3cm},
  bn/.style={draw,thick,rounded corners,fill=blue!05,inner sep=.3cm},
  greenn/.style={draw,thick,rounded corners,fill=green!20,inner sep=.3cm},
  to/.style={
    ->,>=stealth',shorten >=1pt,semithick,font=\sffamily\footnotesize
  },
}

\IfFileExists{upquote.sty}{\usepackage{upquote}}{}
\begin{document}

\title{A computer algebra system for \pl{R}: \pl{Macaulay2} and the \pkg{m2r} package}

\author[1]{David Kahle\thanks{\href{mailto:david.kahle@gmail.com}{david.kahle@gmail.com}}}
\author[2]{Christopher O'Neill\thanks{\href{mailto:coneill@math.ucdavis.edu}{coneill@math.ucdavis.edu}}}
\author[3]{Jeff Sommars\thanks{\href{mailto:sommars1@uic.edu}{sommars1@uic.edu}}}
\affil[1]{Department of Statistical Science, Baylor University}
\affil[2]{Department of Mathematics, University of California, Davis}
\affil[3]{Department of Mathematics, Statistics, and Computer Science, University of Illinois at Chicago}
\date{}

\maketitle




\begin{abstract}
Algebraic methods have a long history in statistics.  The most
prominent manifestation of modern algebra in statistics can be
seen in the field of algebraic statistics, which brings tools
from commutative algebra and algebraic geometry to bear on
statistical problems.  Now over two decades old, algebraic
statistics has applications in a wide range of theoretical and
applied statistical domains.  Nevertheless, algebraic
statistical methods are still not mainstream, mostly due to a
lack of easy off-the-shelf implementations.  In this article
we debut \pkg{m2r}, an \pl{R} package that connects \pl{R} to
\pl{Macaulay2} through a persistent back-end socket connection
running locally or on a cloud server. Topics range from basic
use of \pkg{m2r} to applications and design philosophy. 
\end{abstract}


\section{Introduction}\label{sec:intro}

Algebra, a branch of mathematics concerned with abstraction,
structure, and symmetry, has a long history of applications in
statistics.  For example, Pearson's early work on method of moments
estimation in mixture models ultimately involved systems of polynomial
equations that he painstakingly and remarkably solved by hand
\citep{pearson1894contributions, amendola2016}.  Fisher's work in
design was strongly algebraic and combinatorial, focusing on topics
such as Latin squares \citep{fisher1934statistical}. Invariance and
equivariance continue to form a major pillar of mathematical
statistics through the lens of location-scale families
\citep{pitman1939tests, bondesson1983, lehmanntsh}.  

Apart from the obvious applications of linear algebra, the most
visible manifestations of modern algebra in statistics are found in
the young field of algebraic statistics. Algebraic statistics is
defined broadly as the application of commutative algebra and
algebraic geometry to statistical problems, generally understood to
include applications of other mathematical fields that have
substantial overlap with commutative algebra and algebraic geometry,
such as combinatorics, polyhedral geometry, graph theory, and others
\citep{drton2009lectures, sturmfels1996}. Now a quarter century old,
algebraic statistics has revealed that many statistical areas are
profitably amenable to algebraic investigation, including discrete
multivariate analysis, discrete and Gaussian graphical models,
statistical disclosure limitation, phylogenetics, Bayesian statistics,
and more. Nevertheless, while the field is well-established and
actively growing, advances in algebraic statistical methods are still
not mainstream among applied statisticians, largely due to the lack of
off-the-shelf implementations of key algebraic algorithms in
mainstream statistical software.  In this article we debut \pkg{m2r},
a key piece to the puzzle of applied algebraic statistics in \pl{R}.


\subsection{\pl{Macaulay2} and the \pkg{m2r} \pl{R} package}

\pl{Macaulay2} is a state-of-the-art, open-source computer algebra
system designed to perform computations in commutative algebra and
algebraic geometry \citep{M2}.  More than twenty years old, the
software has a large code base with many community members actively
developing add-on packages. In addition, \pl{Macaulay2} links to other
major open source software in the mathematics community, such as
\pl{Normaliz} \citep{normaliz, normaliz2, normalizmacaulay2},
\pl{4ti2} \citep{4ti2}, and \pl{PHCpack} \citep{phcpack,
phcpackmacaulay2}, through a variety of interfaces.  Natively,
\pl{Macaulay2} is well-known for its efficiency with large algebraic
computations, among other things. 

One of the primary benefits of \pl{Macaulay2} is its efficiency with
large algebraic computations. For instance, Gr\"{o}bner basis
computations comprise the core of many algorithms central to
computational algebra. Some of these computations take many hours and
produce output consisting of several thousand polynomials or
polynomials with several thousand terms. Often, the \pl{Macaulay2}
user will not be interested in the entire output, but only certain
properties; \pl{Macaulay2} allows the user to specify relevant
properties to return, such as the dimension of the solution set or the
highest degree term that appears.  

\pl{R} is increasingly the programming lingua franca of the statistics
community, but it has very limited native support for symbolic
computing  \citep{R}.  \pkg{rSymPy} attempts to alleviate this problem
by connecting \pl{R} to \pl{Python}'s \pkg{SymPy} library
\citep{sympy, rsympy}.  \pkg{mpoly} provides a basic collection of
\pl{R} data structures and methods for multivariate polynomials and
was designed to lay the foundation for a more robust computer algebra
system in \pl{R} \citep{mpoly}.  Unfortunately, neither of these
wholly meet the computational needs of those in the algebraic
statistics community, because neither of them were designed for that
purpose.  Consequently, for years those using algebraic statistical
methods have been forced to go outside of \pl{R} to manually run key
algebraic computations in software such as \pl{Macaulay2} and then
pull the results back into \pl{R}. This error prone and tedious
process is simply one barrier to entry to using algebraic statistics
in \pl{R}. The problem is compounded by users needing to install
\pl{Macaulay2}, which is not cross-platform, and be familiar with the
\pl{Macaulay2} language, which is syntactically and semantically very
different from \pl{R}.

In this article we present the \pkg{m2r} package, which is intended to
help fill this void.  \pkg{m2r} was created at the American
Mathematical Society's 2016 Mathematics Research Community gathering
on algebraic statistics. It connects \pl{R} to a persistent local or
remote \pl{Macaulay2} session and leverages \pkg{mpoly}'s existing
infrastructure to provide wrappers for commonly used algebraic
algorithms in a way that naturally fits into the \pl{R} ecosystem,
alleviating the need to learn \pl{Macaulay2}.  It is our hope that
\pkg{m2r} will provide a flexible framework for computations in the
algebraic statistics community and beyond.

The outline of the article is as follows. In Section~\ref{sec:theory}
we provide a basic overview of the relevant algebraic and geometric
concepts used in the rest of the article; we also provide references
to learn more.  In Section~\ref{sec:basic-usage} we present a basic
demo of \pkg{m2r} to get up and running.
Section~\ref{sec:applications} follows with two applications of
interest to \pl{R} users: using \pkg{m2r} to exactly solve systems of
nonlinear algebraic equations and applying \pkg{m2r} to better
understand conditional independence models on multiway contingency
tables.  Next, Sections~\ref{sec:internals} and~\ref{sec:connecting}
provide an overview of how \pkg{m2r} works internally, first by
describing the design philosophy and then by demonstrating how
\pkg{m2r} connects \pl{R} to \pl{Macaulay2}, which need not be
installed locally on the user's machine.  We conclude with a brief
discussion of future directions in Section~\ref{sec:discussion}.

\section{Theory and applications}\label{sec:theory}

In this section we provide a basic introduction to the algebraic and
geometric objects described in the remainder of this work.  We aim for
understandability over precision, and so in some cases bend the truth
a bit.  There are accessible texts for more precise definitions; we
direct the reader to \cite{gallian} for the basics of modern algebra,
and \cite{cox} for the basics of commutative algebra and
algebraic~geometry. 

Broadly speaking, the mathematical discipline of algebra deals with
sets of objects with certain well-defined operations between their
elements that result in other elements of the set (e.g.\ the sum of
two numbers is a number). At a basic level, modern algebra has focused
on three such objects, in order of increasing structure: groups,
rings, and fields.  A \defn{group} is a set along with a single binary
operation ``$+$'' in which every element has an inverse.  For example,
the integers ($\Z$) form a group; $0$ is the \defn{identity} element
($x + 0 = x$ for any $x \in \Z$) and the inverse of any integer is its
negative ($x + (-x) = 0$).  A \defn{ring} is a group with a second
operation ``$\cdot$'' under which elements need not have inverses.
For example, $\Z$ is also a ring; the product of two integers is an
integer, and the multiplicative identity is the number $1$ ($1 \cdot x
= x$ for any $x \in \Z$), but $2$ has no multiplicative inverse since
$1/2$ is not an integer.  A \defn{field} is a ring with multiplicative
inverses, i.e.~a ring where division is defined.  As such, the
integers form a ring but not a field.  On the other hand, the rational
numbers $\Q$ do form a field, as do the real numbers $\R$ and the
complex numbers $\C$.  Throughout this paper, all group and ring
operations will be \defn{commutative}, or order invariant, e.g.\ $5
\cdot 2 = 2 \cdot 5$.  

Among each class of objects, special subsets are distinguished.  For
example, a \defn{subgroup} of a group is a subset of a group that is
itself a group, e.g.\ the even integers.  The field of commutative
algebra focuses on commutative rings and distinguished subsets called
ideals.  An \defn{ideal} is a subgroup of a ring that ``absorbs''
elements of the ring under multiplication.  For example, the even
integers $\mc{I} \subset \Z$ are an ideal of the ring of integers;
$\mc{I}$ is a group under addition, and if you multiply an even number
by any integer, the result is even and thus in $\mc{I}$.  Note that
ideals are not necessarily rings, as they usually do not contain the
multiplicative identity $1$ (in fact, any ideal containing $1$ must
contain every element of the ring).  Special supersets are also
distinguished. For example a \defn{field extension} $\F'$ of a field
$\F$ is a superset of $\F$ that is a field under the same operations
as $\F$, e.g. $\C$ and $\R$ of $\Q$.

As mathematical objects, the set of polynomials in one or several
variables forms a commutative ring.  Since the general multivariate
setting is as accessible as the more familiar univariate setting, we
go straight to multivariate polynomials.  Let $\ve{x}$ denote an
$n$-tuple $\ve{x} = (x_{1}, x_{2}, \ldots, x_{n})$ of variables.  A
\defn{monomial} is a product of the variables of the form
\begin{equation}
\ve{x}^{\veg{\al}} 
\ \ = \ \ x_{1}^{\al_{1}} x_{2}^{\al_{2}} \cdots x_{n}^{\al_{n}},
\quad \al_{i} \in \N_{0} = \set{0, 1, 2, \ldots}.
\end{equation}
A \defn{polynomial} $f$ is a finite linear combination of monomials
whose coefficients are drawn from some ring $\K$ (often a field such
as $\Q$, $\R$, or $\C$).  The set of all polynomials with coefficients
in $\K$ is denoted $\K[\ve{x}]$.  For example, $f(x, y) = 3x - 2y \in
\Z[x,y]$.  Obviously, adding, subtracting, and multiplying polynomials
results in another polynomial after simplification.

One way to create an ideal in a polynomial ring is simply to generate
one from a collection of polynomials.  If $f_{1}, \ldots, f_{m}$ is a
collection of $m$ polynomials in $\K[\ve{x}]$, the \defn{ideal
generated by $f_{1}, \ldots, f_{m}$} is the set
\begin{equation}
\langle f_{1}, \ldots, f_{m} \rangle 
\ \ = \ \ \set{r_{1}f_{1} + \cdots + r_{m}f_{m} : r_{k} \in \K[\ve{x}] \ \mbox{for} \ k = 1, \ldots, m}
\ \ \subseteq \ \ \K[\ve{x}].
\end{equation}
In particular, this set is the smallest ideal containing $f_{1},
\ldots, f_{m}$.  The generating polynomials $f_{1}, \ldots, f_{m}$ are
called a \defn{basis} of the ideal.  Obviously, ideals are infinitely
large collections of polynomials.  However, they typically aren't
\emph{all} polynomials; in the ring $\Z[x,y]$, $\mc{I} = \langle x, y
\rangle$ is an ideal, and $\Z[x,y] \setminus \mc{I}$ consists of all
polynomials with nonzero constant term.  A remarkable result known as
the Hilbert basis theorem states that every ideal has a finite
generating set, i.e.\ a finite basis.  However, bases need not be
unique.  \defn{Gr\"{o}bner bases} are generating sets with some
additional structure and are central objects in computational
commutative algebra.  In~general, it can be difficult to answer
questions such as whether or not two ideals are equal, or if a
particular polynomial is contained in an ideal.  If one has a
Gr\"{o}bner basis however, these questions can be answered relatively
easily.  

There are a number of algorithms known to convert a given collection
of polynomials $f_{1}, \ldots, f_{m}$ into a Gr\"{o}bner basis $g_{1},
\ldots, g_{m'}$.  The first historically and simplest is Buchberger's
algorithm, and all major computer algebra systems implement a variant
of it, including \pl{Macaulay2} and \pl{Singular} \citep{buchberger,
M2, sing}. Optimizing Gr\"{o}bner basis computations continues to be
an active area of research in computational algebraic geometry, and
the aforementioned software packages are regularly updated with newer
and faster implementations.  

Algebraic geometry is the field of mathematics interested in
understanding the geometric structure of zero sets of polynomials,
called \defn{varieties} or \defn{algebraic sets}.  Concretely, the
\defn{variety generated by $f_{1}, \ldots, f_{m}$} is the set of
vectors $\ve{x} \in \K^{n}$ where all the polynomials evaluate to
zero.
\begin{equation}
\varty{f_{1}, \ldots, f_{m}}
\ \ = \ \ \set{\ve{x} \in \K^{n} : f_{1}(\ve{x}) = \cdots = f_{m}(\ve{x}) = 0}.
\end{equation}
Sometimes a field extension of $\K$ is used instead of $\K$ so that,
for example, we could consider the set of solutions in $\R^{n}$ of a
polynomial with coefficients in $\Z$ (which are of course also in
$\R$).  Varieties are geometric objects.  For example, the variety
generated by the polynomial $x^{2} + y^{2} - 1 \in \R[x,y]$ is the
unit circle; it consists of all pairs $(x,y) \in \R^{2}$ such that
$x^{2} + y^{2} = 1$.
 
A system of polynomial equations can be converted into a collection of
polynomials by moving every term to one side, leaving the other side
to be just zeros; this is a common technique in algebraic geometry.
The variety of the resulting set of polynomials is the set of common
solutions to the original list of equations.  If no solutions exist,
the system is said to be \defn{inconsistent}; if there are a finite
number of solutions, the variety is said to be \defn{zero
dimensional}; and if there are an infinite number of solutions, the
variety is said to be \defn{positive dimensional}.

Note that this construction is a nonlinear generalization of linear
algebra.  Linear algebra studies polynomials of degree one, where
every term has at most one variable and its exponent is one.  The
varieties are linear varieties: the empty set, a single point, lines,
planes, or hyperplanes.  By contrast, in general varieties can be
significantly more complicated.  They can be curved, come to sharp
points, be self intersecting, or even disconnected.  Unions of
varieties are varieties by multiplying their generating sets pairwise,
and intersections of varieties are varieties by simply taking all the
generators of both. Consequently, given a variety $V$ it make sense to
talk about its \defn{minimal decomposition}, the representation of $V$
as a union $V = \bigcup V_{i}$ of smaller \defn{irreducible varieties}
$V_{i}$ that can not be further decomposed (i.e. if $V_{i} = W_{1}
\cup W_{2}$ for varieties $W_{1}$ and $W_{2}$, either $V_{i} = W_{1}$
or $V_{i} = W_{2}$). Such unions are always finite. The
\defn{dimension of a variety} is the maximum dimension of its
irreducible components, which are in turn defined as the dimension of
a tangent hyperplane at a generic point, e.g. the dimension of the
circle is 1 since (tangent) lines are one dimensional.

There is a rich interplay between polynomial ideals and varieties that
forms the core of algebraic geometry and allows us to align geometric
structures and procedures with algebraic ones in a near one-to-one
fashion.  In this setting, Gr\"{o}bner bases play a major role.  If
$\mc{I}$ is an ideal, the variety of $\mc{I}$, $\varty{\mc{I}}$, is
the zero set of all the polynomials in $\mc{I}$.  If $\mc{I}$ is
generated by the polynomials $f_{1}, \ldots, f_{m}$, then
$\varty{\mc{I}} = \varty{f_{1}, \ldots, f_{m}}$; in particular,
different bases of ideals generate identical varieties.  In algebraic
geometry, Gr\"{o}bner bases are good choices for bases for myriad
reasons.  For example, if the variety $\varty{\mc{I}}$ is zero
dimensional, a (lexicographic) Gr\"{o}bner basis is structured in such
a way that the equations can be solved one at a time and
back-substituted into the others, much in the same way that in a
linear system with a unique solution, after Gaussian elimination
solutions can be read off and back-substituted one by one.  Many
geometric properties of varieties, such as their dimension or an
irreducible decomposition, can also be easily computed using
Gr\"{o}bner bases.

\section{Basic usage}\label{sec:basic-usage}

This section showcases the basic capabilities of \pkg{m2r} and some of
the ways that \pl{Macaulay2} can be used. 


\subsection{Loading \pkg{m2r}}\label{sec:loading}

\pkg{m2r} is loaded like any other \pl{R} package:
\begin{knitrout}
\definecolor{shadecolor}{rgb}{0.969, 0.969, 0.969}\color{fgcolor}\begin{kframe}
\begin{alltt}
\hlstd{R> }\hlkwd{library}\hlstd{(m2r)}
\end{alltt}

{\ttfamily\noindent\itshape\color{messagecolor}{Loading required package: mpoly}}

{\ttfamily\noindent\itshape\color{messagecolor}{Loading required package: stringr}}

{\ttfamily\noindent\color{warningcolor}{Warning: package 'stringr' was built under R version 3.3.2}}

{\ttfamily\noindent\itshape\color{messagecolor}{M2 found in /usr/local/macaulay2/bin}}\end{kframe}
\end{knitrout}
\noindent The first two lines of output indicate that \pkg{m2r}
depends on \pkg{mpoly} and \pkg{stringr}. The packages \pkg{mpoly} and
\pkg{stringr} manipulate and store multivariate polynomials and
strings, respectively \citep{mpoly, stringr}.  The third line
indicates that \verb|M2|, the \pl{Macaulay2} executable, was found on
the user's machine at the given path, and that the version of
\pl{Macaulay2} in that directory will be used for computations. When
loaded on a Unix-like machine, \pkg{m2r} looks for \verb|M2| on the
user's machine by searching through \verb|~/.bash_profile|, or if
nonexistent, \verb|~/.bashrc| and \verb|~/.profile|. \pkg{m2r} stores
the first place  \verb|M2|is found in the option
{\tt \hlstd{m2r}\hlopt{\$}\hlstd{m2\_path}}.\footnote{Note that \pkg{m2r} will not
necessarily use whatever is on the user's typical \code{PATH} variable
because when \pl{R} makes {\tt \hlkwd{system}\hlstd{()}} calls, it does not load
the user's personal configuration files.  If a different path is
desired, the user can easily change this option with the function
{\tt \hlkwd{set\_m2\_path}\hlstd{()}}.} 

When \pkg{m2r} is loaded, \pl{Macaulay2} is searched for but not
initialized.  The actual initialization and subsequent connection to
\pl{Macaulay2} by \pkg{m2r} takes place when \pl{R} first calls a
\pl{Macaulay2} function through \pkg{m2r}.


\subsection{\pkg{m2r} basics}\label{sec:basics}

The basic interface to \pl{Macaulay2} is provided by the
{\tt \hlkwd{m2}\hlstd{()}} function. {\tt \hlkwd{m2}\hlstd{()}} accepts a character string
containing \pl{Macaulay2} code, sends it to \pl{Macaulay2} to be
evaluated, and brings the output back into \pl{R}. For example, like
all computer algebra systems, \pl{Macaulay2} supports basic
arithmetic:
\begin{knitrout}
\definecolor{shadecolor}{rgb}{0.969, 0.969, 0.969}\color{fgcolor}\begin{kframe}
\begin{alltt}
\hlstd{R> }\hlkwd{m2}\hlstd{(}\hlstr{"1 + 1"}\hlstd{)}
\end{alltt}

{\ttfamily\noindent\itshape\color{messagecolor}{Starting M2... }}

{\ttfamily\noindent\itshape\color{messagecolor}{done.}}\begin{verbatim}
[1] "2"
\end{verbatim}
\end{kframe}
\end{knitrout}
\noindent Unlike most \pkg{m2r} functions, {\tt \hlkwd{m2}\hlstd{()}} does not
parse the \pl{Macaulay2} output into an \pl{R} data structure.  This
can be seen in the result above being a character and not a numeric,
but it is even more evident when evaluating a floating point number:
\begin{knitrout}
\definecolor{shadecolor}{rgb}{0.969, 0.969, 0.969}\color{fgcolor}\begin{kframe}
\begin{alltt}
\hlstd{R> }\hlkwd{m2}\hlstd{(}\hlstr{"1.2"}\hlstd{)}
\end{alltt}
\begin{verbatim}
[1] ".12p53e1"
\end{verbatim}
\end{kframe}
\end{knitrout}
\noindent Parsing the output is a delicate task accomplished by the
{\tt \hlkwd{m2\_parse}\hlstd{()}} function:
\begin{knitrout}
\definecolor{shadecolor}{rgb}{0.969, 0.969, 0.969}\color{fgcolor}\begin{kframe}
\begin{alltt}
\hlstd{R> }\hlkwd{m2_parse}\hlstd{(}\hlkwd{m2}\hlstd{(}\hlstr{"1.2"}\hlstd{))}
\end{alltt}
\begin{verbatim}
[1] 1.2
\end{verbatim}
\end{kframe}
\end{knitrout}
\noindent We expand on how {\tt \hlkwd{m2\_parse}\hlstd{()}} works as a general
\pl{Macaulay2} parser in Section~\ref{sec:internals}.

One of the great advantages to \pkg{m2r}'s implementation is that it
provides a persistent connection to a \pl{Macaulay2} session running
in the background.  In early versions of \pkg{algstat}, \pl{Macaulay2}
was accessible from \pl{R} through intermediate script files;
\pkg{algstat} saved user supplied \pl{Macaulay2} code to a temporary
file, called \pl{Macaulay2} in script mode to evaluate it, saved the
output to another temporary file, and parsed the output back into
\pl{R} \citep{algstat}. One of the major limitations of this scheme is
that every computation and every variable created on the
\pl{Macaulay2} side is lost once the call is complete.  Unlike
\pkg{algstat}, \pkg{m2r} allows for this kind of persistent connection
to a \pl{Macaulay2} session, which is easy to demonstrate:
\begin{knitrout}
\definecolor{shadecolor}{rgb}{0.969, 0.969, 0.969}\color{fgcolor}\begin{kframe}
\begin{alltt}
\hlstd{R> }\hlkwd{m2}\hlstd{(}\hlstr{"a = 1"}\hlstd{)}
\end{alltt}
\begin{verbatim}
[1] "1"
\end{verbatim}
\begin{alltt}
\hlstd{R> }\hlkwd{m2}\hlstd{(}\hlstr{"a"}\hlstd{)}
\end{alltt}
\begin{verbatim}
[1] "1"
\end{verbatim}
\end{kframe}
\end{knitrout}
\noindent When not actively running code, the \pl{Macaulay2} session
sits, listening for commands issued by \pl{R}.  The details of the
connection are described in detail in Section~\ref{sec:connecting}.  

While the \pl{Macaulay2} session is live, it helps to have \pl{R}-side
functions that access it in a natural way.  Because of this, just as
there are functions such as {\tt \hlkwd{ls}\hlstd{()}} and {\tt \hlkwd{exists}\hlstd{()}} in
\pl{R}, \pkg{m2r} provides analogues for the background \pl{Macaulay2}
session:

\begin{knitrout}
\definecolor{shadecolor}{rgb}{0.969, 0.969, 0.969}\color{fgcolor}\begin{kframe}
\begin{alltt}
\hlstd{R> }\hlkwd{m2_ls}\hlstd{()}
\end{alltt}
\begin{verbatim}
[1] "a"
\end{verbatim}
\begin{alltt}
\hlstd{R> }\hlkwd{m2_exists}\hlstd{(}\hlkwd{c}\hlstd{(}\hlstr{"a"}\hlstd{,} \hlstr{"b"}\hlstd{))}
\end{alltt}
\begin{verbatim}
[1]  TRUE FALSE
\end{verbatim}
\begin{alltt}
\hlstd{R> }\hlkwd{m2_getwd}\hlstd{()}
\end{alltt}
\begin{verbatim}
[1] "/Users/david_kahle"
\end{verbatim}
\end{kframe}
\end{knitrout}
\noindent {\tt \hlkwd{m2\_ls}\hlstd{()}} also accepts the argument
{\tt \hlstd{all.names} \hlkwb{=} \hlnum{TRUE}}, which gives a larger listing of the
variables defined in the \pl{Macaulay2} session, much like
{\tt \hlkwd{ls}\hlstd{(}\hlkwc{all.names} \hlstd{=} \hlnum{TRUE}\hlstd{)}}. These additional variables fall into
two categories: output variables returned by \pl{Macaulay2} and
\pkg{m2r} variables used to manage the connection. In \pl{Macaulay2},
the output of each executed line of code is stored as a variable bound
to the symbol \code{o} followed by the line number executed.  For
example, the output of the first executed line is \code{o1}. These are
accessible through \pkg{m2r} as, for example, \code{m2o1}; however,
since \pkg{m2r}'s internal connection itself makes calls to
\pl{Macaulay2}, the numbering is somewhat unpredictable.  This is why
they don't show up in {\tt \hlkwd{m2\_ls}\hlstd{()}} by default.  The internal
variables that \pkg{m2r} uses to manage the persistent connection to
\pl{Macaulay2} are called \verb|m2rint*| and generally shouldn't be
accessed by the user; we provide more on this in
Section~\ref{sec:structures}.

%


\subsection{Commutative algebra and algebraic geometry}\label{sec:comalg}

\pl{Macaulay2} is designed for computations in commutative algebra and
algebraic geometry.  Consequently, algebraic structures such as
polynomial rings and ideals are of primary interest.  While the
{\tt \hlkwd{m2}\hlstd{()}} function suffices at a basic level for these kinds of
operations in \pl{R}, \pkg{m2r} provides a number of wrapper functions
and data structures that facilitate interacting with \pl{Macaulay2} in
a way that is significantly more familiar to \pl{R} users.  In the
remainder of this section we showcase these kinds of functions in
action.  We begin with rings and ideals, the basic algebraic
structures in commutative algebra, and the computation of Gr\"{o}bner
bases.

Polynomial rings can be created with the {\tt \hlkwd{ring}\hlstd{()}} function:
\begin{knitrout}
\definecolor{shadecolor}{rgb}{0.969, 0.969, 0.969}\color{fgcolor}\begin{kframe}
\begin{alltt}
\hlstd{R> }\hlstd{(R} \hlkwb{<-} \hlkwd{ring}\hlstd{(}\hlstr{"t"}\hlstd{,} \hlstr{"x"}\hlstd{,} \hlstr{"y"}\hlstd{,} \hlstr{"z"}\hlstd{,} \hlkwc{coefring} \hlstd{=} \hlstr{"QQ"}\hlstd{))}
\end{alltt}
\begin{verbatim}
M2 Ring: QQ[t,x,y,z], grevlex order
\end{verbatim}
\end{kframe}
\end{knitrout}
\noindent As described in Section~\ref{sec:theory}, polynomial rings
are comprised of two basic components: a collection of variables, and
a coefficient ring, often a field.  In \pl{Macaulay2}, several special
key words exist that refer to commonly used coefficient rings: the
integers $\Z$ (\code{ZZ}), the rational numbers $\Q$ (\code{QQ}), the
real numbers $\R$ (\code{RR}), and the complex numbers $\C$
(\code{CC}).  Polynomial rings and related algorithms often benefit
from total orders on their monomials.  These can be supplied through
{\tt \hlkwd{ring}\hlstd{()}}'s {\tt \hlstd{order}} argument, which by default sets
{\tt \hlstd{order} \hlkwb{=} \hlstr{"grevlex"}}, the graded reverse lexicographic order.

Ideals of rings can be specified with the {\tt \hlkwd{ideal}\hlstd{()}} function
as follows:
\begin{knitrout}
\definecolor{shadecolor}{rgb}{0.969, 0.969, 0.969}\color{fgcolor}\begin{kframe}
\begin{alltt}
\hlstd{R> }\hlstd{(I} \hlkwb{<-} \hlkwd{ideal}\hlstd{(}\hlstr{"t^4 - x"}\hlstd{,} \hlstr{"t^3 - y"}\hlstd{,} \hlstr{"t^2 - z"}\hlstd{))}
\end{alltt}
\begin{verbatim}
M2 Ideal of ring QQ[t,x,y,z] (grevlex) with generators : 
< t^4  -  x,  t^3  -  y,  t^2  -  z >
\end{verbatim}
\end{kframe}
\end{knitrout}
\noindent They are defined relative to the last ring used that
contains all the variables referenced.  If no such ring exists, you
get an error. A common mistake along these lines is to try to
reference a variable that cannot be scoped to a previously defined
ring:
\begin{knitrout}
\definecolor{shadecolor}{rgb}{0.969, 0.969, 0.969}\color{fgcolor}\begin{kframe}
\begin{alltt}
\hlstd{R> }\hlkwd{m2}\hlstd{(}\hlstr{"u + 1"}\hlstd{)}
\end{alltt}

{\ttfamily\noindent\bfseries\color{errorcolor}{Error: Macaulay2 Error!}}\end{kframe}
\end{knitrout}
\noindent In a situation where several rings are have been used, the
{\tt \hlkwd{use\_ring}\hlstd{()}} function is helpful to specify which specific
ring to use.  For example, {\tt \hlkwd{use\_ring}\hlstd{(R)}}.

Gr\"{o}bner bases of ideals are computed with {\tt \hlkwd{gb}\hlstd{()}}:
\begin{knitrout}
\definecolor{shadecolor}{rgb}{0.969, 0.969, 0.969}\color{fgcolor}\begin{kframe}
\begin{alltt}
\hlstd{R> }\hlkwd{gb}\hlstd{(I)}
\end{alltt}
\begin{verbatim}
z^2  -  x
z t  -  y
-1 z x  +  y^2
-1 x  +  t y
-1 z y  +  x t
-1 z  +  t^2
\end{verbatim}
\end{kframe}
\end{knitrout}
\noindent  To provide a more natural feel, {\tt \hlkwd{ideal}\hlstd{()}} and
{\tt \hlkwd{gb}\hlstd{()}} are overloaded to accept any of many types of input,
including \code{mpoly} and \code{mpolyList} objects.  For example,
instead of {\tt \hlkwd{gb}\hlstd{()}} working on an ideal object, it can work
directly on a collection of polynomials:
\begin{knitrout}
\definecolor{shadecolor}{rgb}{0.969, 0.969, 0.969}\color{fgcolor}\begin{kframe}
\begin{alltt}
\hlstd{R> }\hlkwd{gb}\hlstd{(}\hlstr{"t^4 - x"}\hlstd{,} \hlstr{"t^3 - y"}\hlstd{,} \hlstr{"t^2 - z"}\hlstd{)}
\end{alltt}
\begin{verbatim}
z^2  -  x
z t  -  y
-1 z x  +  y^2
-1 x  +  t y
-1 z y  +  x t
-1 z  +  t^2
\end{verbatim}
\end{kframe}
\end{knitrout}

You may have noticed something strange in this last call:
{\tt \hlkwd{gb}\hlstd{(I)}} only took one argument, whereas {\tt \hlkwd{gb}\hlstd{(}\hlstr{"t\string^4 - x"}\hlstd{,}}{\tt \hlstr{"t\string^3 - y"}\hlstd{,} \hlstr{"t\string^2 - z"}\hlstd{)}} took three, but they performed the same task.
This is possible because of nonstandard evaluation in \pl{R}
\citep{wickham2014advanced, standardnonstandard}.  While nonstandard
evaluation is very convenient, it does have drawbacks.  In particular,
it tends to be hard to use functions that use nonstandard evaluation
inside other functions, so using {\tt \hlkwd{gb}\hlstd{()}}, for example, inside a
function in a package that depends on \pkg{m2r} can be tricky.  To
alleviate this problem, each of {\tt \hlkwd{ring}\hlstd{()}}, {\tt \hlkwd{ideal}\hlstd{()}},
and {\tt \hlkwd{gb}\hlstd{()}} has a standard evaluation version that tends to be
easier to program with and incorporate into packages .  Following the
\pkg{dplyr}/\pkg{tidyverse} naming convention \citep{dplyr}, these
functions have the same name followed by an underscore:
{\tt \hlkwd{ring\_}\hlstd{()}}, {\tt \hlkwd{ideal\_}\hlstd{()}}, and {\tt \hlkwd{gb\_}\hlstd{()}}.  To see
the difference between standard and nonstandard evaluation, compare
the previous {\tt \hlkwd{gb}\hlstd{()}} call, which depends on nonstandard
evaluation, to this call to {\tt \hlkwd{gb\_}\hlstd{()}}, which uses standard
evaluation:
\begin{knitrout}
\definecolor{shadecolor}{rgb}{0.969, 0.969, 0.969}\color{fgcolor}\begin{kframe}
\begin{alltt}
\hlstd{R> }\hlstd{polys} \hlkwb{<-} \hlkwd{c}\hlstd{(}\hlstr{"t^4 - x"}\hlstd{,} \hlstr{"t^3 - y"}\hlstd{,} \hlstr{"t^2 - z"}\hlstd{)}
\hlstd{R> }\hlkwd{gb_}\hlstd{(polys,} \hlkwc{ring} \hlstd{= R)}
\end{alltt}
\begin{verbatim}
z^2  -  x
z t  -  y
-1 z x  +  y^2
-1 x  +  t y
-1 z y  +  x t
-1 z  +  t^2
\end{verbatim}
\end{kframe}
\end{knitrout}
\noindent Though the distinction is not as obvious, {\tt \hlkwd{gb}\hlstd{(I)}}
and {\tt \hlkwd{gb\_}\hlstd{(I)}} both work and result in the same computation.
The latter, however, is more appropriate for use inside packages.

Radicals of ideals, which can be thought of as a method of eliminating
root multiplicity, can be computed with {\tt \hlkwd{radical}\hlstd{()}}.  We note
that \pl{Macaulay2} has only implemented this feature for polynomial
rings over the rationals $\Q$ (\code{QQ}) and finite fields $\Z/p$
(\code{ZZ/p}).
\begin{knitrout}
\definecolor{shadecolor}{rgb}{0.969, 0.969, 0.969}\color{fgcolor}\begin{kframe}
\begin{alltt}
\hlstd{R> }\hlkwd{ring}\hlstd{(}\hlstr{"x"}\hlstd{,} \hlkwc{coefring} \hlstd{=} \hlstr{"QQ"}\hlstd{)}
\end{alltt}
\begin{verbatim}
M2 Ring: QQ[x], grevlex order
\end{verbatim}
\begin{alltt}
\hlstd{R> }\hlstd{I} \hlkwb{<-} \hlkwd{ideal}\hlstd{(}\hlstr{"x^2"}\hlstd{)}
\hlstd{R> }\hlkwd{radical}\hlstd{(I)}
\end{alltt}
\begin{verbatim}
M2 Ideal of ring QQ[x] (grevlex) with generator : 
< x >
\end{verbatim}
\end{kframe}
\end{knitrout}

\defn{Ideal saturation} is a more complex process than the scope of
this work entails, but it is worth mentioning as it has a variety of
applications.  Loosely speaking, the saturation of an ideal $\mc{I}$
by another ideal $\mc{J}$, denoted $\mc{I}:\mc{J}^{\infty}$, is an
ideal containing $\mc{I}$ and any additional polynomials obtained by
``dividing out'' elements of $\mc{J}$.
Enlarging an ideal reduces the size of its corresponding variety; more
polynomials means more conditions a point $\ve{x} \in \K^{n}$ in the
variety must satisfy.  On the variety side, saturation is intended to
remove components of the variety that are known to be nonzero. 
In \pkg{m2r}, saturation can be computed with {\tt \hlkwd{saturate}\hlstd{()}}.
Notice in what follows saturation of the ideal $\langle (x-1)x(x+1)
\rangle$, with variety $-1$, $0$, and $1$, by the ideal $\langle x
\rangle$ removes the solution $x = 0$:
\begin{knitrout}
\definecolor{shadecolor}{rgb}{0.969, 0.969, 0.969}\color{fgcolor}\begin{kframe}
\begin{alltt}
\hlstd{R> }\hlstd{I} \hlkwb{<-} \hlkwd{ideal}\hlstd{(}\hlstr{"(x-1) x (x+1)"}\hlstd{)}
\hlstd{R> }\hlstd{J} \hlkwb{<-} \hlkwd{ideal}\hlstd{(}\hlstr{"x"}\hlstd{)}
\hlstd{R> }\hlkwd{saturate}\hlstd{(I, J)}
\end{alltt}
\begin{verbatim}
M2 Ideal of ring QQ[x] (grevlex) with generator : 
< x^2  -  1 >
\end{verbatim}
\end{kframe}
\end{knitrout}
\noindent The closely related concept of an \defn{ideal quotient}
$\mc{I}:\mc{J}$ can be computed with {\tt \hlkwd{quotient}\hlstd{()}}.

The \defn{primary decomposition} of an ideal is the algebraic analogue
of the minimal decomposition of a variety into irreducible components.
Primary decompositions can be computed with
{\tt \hlkwd{primary\_decomposition}\hlstd{()}}.  The result is a list of ideals
(class {\tt \hlstd{m2\_ideal\_list}}).  For example, the ideal $\langle xz,
yz \rangle$ corresponds to the variety that is the union of the
$xy$-plane and the $z$ axis.  That notion can be recaptured with
primary decomposition:
\begin{knitrout}
\definecolor{shadecolor}{rgb}{0.969, 0.969, 0.969}\color{fgcolor}\begin{kframe}
\begin{alltt}
\hlstd{R> }\hlkwd{use_ring}\hlstd{(R)}
\hlstd{R> }\hlstd{I} \hlkwb{<-} \hlkwd{ideal}\hlstd{(}\hlstr{"x z"}\hlstd{,} \hlstr{"y z"}\hlstd{)}
\hlstd{R> }\hlstd{(ideal_list} \hlkwb{<-} \hlkwd{primary_decomposition}\hlstd{(I))}
\end{alltt}
\begin{verbatim}
M2 List of ideals of QQ[t,x,y,z] (grevlex) : 
< z >
< x,  y >
\end{verbatim}
\end{kframe}
\end{knitrout}
\noindent The dimensions of the ideals, which correspond to the
dimensions of their analogous varieties, can be computed with
{\tt \hlkwd{dimension}\hlstd{()}}:
\begin{knitrout}
\definecolor{shadecolor}{rgb}{0.969, 0.969, 0.969}\color{fgcolor}\begin{kframe}
\begin{alltt}
\hlstd{R> }\hlkwd{dimension}\hlstd{(ideal_list)}
\end{alltt}
\begin{verbatim}
M2 List
[[1]]
[1] 3

[[2]]
[1] 2
\end{verbatim}
\end{kframe}
\end{knitrout}

Several other functions exist that aid in whatever one may want to do
with ideals.  For example, sums, products, and equality testing are
all defined as S3 methods of those \pkg{base} functions:
\begin{knitrout}
\definecolor{shadecolor}{rgb}{0.969, 0.969, 0.969}\color{fgcolor}\begin{kframe}
\begin{alltt}
\hlstd{R> }\hlstd{I} \hlkwb{<-} \hlkwd{ideal}\hlstd{(}\hlstr{"x"}\hlstd{,} \hlstr{"y"}\hlstd{)}
\hlstd{R> }\hlstd{J} \hlkwb{<-} \hlkwd{ideal}\hlstd{(}\hlstr{"z"}\hlstd{)}
\hlstd{R> }\hlstd{I} \hlopt{+} \hlstd{J}
\end{alltt}
\begin{verbatim}
M2 Ideal of ring QQ[t,x,y,z] (grevlex) with generators : 
< x,  y,  z >
\end{verbatim}
\begin{alltt}
\hlstd{R> }\hlstd{I} \hlopt{*} \hlstd{J}
\end{alltt}
\begin{verbatim}
M2 Ideal of ring QQ[t,x,y,z] (grevlex) with generators : 
< x z,  z y >
\end{verbatim}
\begin{alltt}
\hlstd{R> }\hlstd{I} \hlopt{==} \hlstd{J}
\end{alltt}
\begin{verbatim}
[1] FALSE
\end{verbatim}
\end{kframe}
\end{knitrout}
\noindent These can be combined with previous functions to great
effect.  For instance, it is simple to script a function to check
whether an ideal is radical:
\begin{knitrout}
\definecolor{shadecolor}{rgb}{0.969, 0.969, 0.969}\color{fgcolor}\begin{kframe}
\begin{alltt}
\hlstd{R> }\hlstd{is.radical} \hlkwb{<-} \hlkwa{function} \hlstd{(}\hlkwc{I}\hlstd{) I} \hlopt{==} \hlkwd{radical}\hlstd{(I)}
\hlstd{R> }\hlkwd{is.radical}\hlstd{(I)}
\end{alltt}
\begin{verbatim}
[1] TRUE
\end{verbatim}
\end{kframe}
\end{knitrout}


In recent years \pkg{magrittr}'s pipe operator \verb|%>%| has become a
mainstream tool in the \pl{R} community, easing the thought process of
programming and clarifying code \citep{magrittr}. The pipe operator
semantically equates the expression {\tt \hlstd{x}} \verb|%>%| {\tt
\hlkwd{f}\hlstd{(y)}} with the more basic \pl{R} expression
{\tt \hlkwd{f}\hlstd{(x,y)}} and the simpler expression {\tt \hlstd{x}} \verb|%>%|
{\tt \hlstd{f}} with {\tt \hlkwd{f}\hlstd{(x)}}. This tool is also very beneficial
in conjunction with \pkg{m2r}. For example, the following code
performs the previous decomposition analysis: it creates an ideal,
decomposes it, and determines the dimension of each component, all in
one simple line of code readable from left to right:
\begin{knitrout}
\definecolor{shadecolor}{rgb}{0.969, 0.969, 0.969}\color{fgcolor}\begin{kframe}
\begin{alltt}
\hlstd{R> }\hlkwd{library}\hlstd{(magrittr)}
\hlstd{R> }\hlkwd{ideal}\hlstd{(}\hlstr{"x z"}\hlstd{,} \hlstr{"y z"}\hlstd{)} \hlopt{%>%}  \hlstd{primary_decomposition} \hlopt{%>%} \hlstd{dimension}
\end{alltt}
\begin{verbatim}
M2 List
[[1]]
[1] 3

[[2]]
[1] 2
\end{verbatim}
\end{kframe}
\end{knitrout}


\subsection{Other examples of \pl{Macaulay2} functionality}\label{sec:other}

In addition to implementations of the basic \pl{Macaulay2} objects and
algorithms of commutative algebra described above, \pkg{m2r} includes
implementations of other algorithms that one might expect in a
computer algebra system.  For example, the prime decomposition of an
integer can be computed with \pkg{m2r}'s {\tt \hlkwd{factor\_n}\hlstd{()}}:
\begin{knitrout}
\definecolor{shadecolor}{rgb}{0.969, 0.969, 0.969}\color{fgcolor}\begin{kframe}
\begin{alltt}
\hlstd{R> }\hlstd{(x} \hlkwb{<-} \hlnum{2}\hlopt{^}\hlnum{5} \hlopt{*} \hlnum{3}\hlopt{^}\hlnum{4} \hlopt{*} \hlnum{5}\hlopt{^}\hlnum{3} \hlopt{*} \hlnum{7}\hlopt{^}\hlnum{2} \hlopt{*} \hlnum{11}\hlopt{^}\hlnum{1}\hlstd{)}
\end{alltt}
\begin{verbatim}
[1] 174636000
\end{verbatim}
\begin{alltt}
\hlstd{R> }\hlstd{(factors} \hlkwb{<-} \hlkwd{factor_n}\hlstd{(x))}
\end{alltt}
\begin{verbatim}
$prime
[1]  2  3  5  7 11

$power
[1] 5 4 3 2 1
\end{verbatim}
\begin{alltt}
\hlstd{R> }\hlkwd{str}\hlstd{(factors)}
\end{alltt}
\begin{verbatim}
List of 2
 $ prime: int [1:5] 2 3 5 7 11
 $ power: int [1:5] 5 4 3 2 1
\end{verbatim}
\begin{alltt}
\hlstd{R> }\hlstd{gmp}\hlopt{::}\hlkwd{factorize}\hlstd{(x)}
\end{alltt}
\begin{verbatim}
Big Integer ('bigz') object of length 15:
 [1] 2  2  2  2  2  3  3  3  3  5  5  5  7  7  11
\end{verbatim}
\end{kframe}
\end{knitrout}
\noindent {\tt \hlkwd{factor\_n}\hlstd{()}} is essentially analogous to \pkg{gmp}'s
{\tt \hlkwd{factorize}\hlstd{()}}, but it is significantly slower due to having to
be passed to \pl{Macaulay2}, computed, passed back, and parsed. On the
other hand, conceptually \pkg{m2r} is factorizing the integer as an
element of a ring, and can do so more generally over other rings, too.
Consequently, polynomials can be factored.  The result is an
\code{mpolyList} object of irreducible polynomials (the analogue to
primes) and a vector of integers, as a list:
\begin{knitrout}
\definecolor{shadecolor}{rgb}{0.969, 0.969, 0.969}\color{fgcolor}\begin{kframe}
\begin{alltt}
\hlstd{R> }\hlkwd{ring}\hlstd{(}\hlstr{"x"}\hlstd{,} \hlstr{"y"}\hlstd{,} \hlkwc{coefring} \hlstd{=} \hlstr{"QQ"}\hlstd{)}
\end{alltt}
\begin{verbatim}
M2 Ring: QQ[x,y], grevlex order
\end{verbatim}
\begin{alltt}
\hlstd{R> }\hlkwd{factor_poly}\hlstd{(}\hlstr{"x^4 - y^4"}\hlstd{)}
\end{alltt}
\begin{verbatim}
$factor
x  -  y
x  +  y
x^2  +  y^2

$power
[1] 1 1 1
\end{verbatim}
\end{kframe}
\end{knitrout}
\noindent One can imagine using this kind of connection, along with
\pl{R}'s random number generators, to experimentally obtain Monte
Carlo answers to a number of mathematical questions.  This kind of
computation has applications in random algebraic geometry and
commutative algebra.

%

A bit more interesting to statisticians may be the implementation of
an algorithm to compute the Smith normal form of a matrix.  The Smith
normal form of a matrix $\ma{M}$ here refers to the decomposition of
an integer matrix $\ma{D} = \ma{P}\ma{M}\ma{Q}$, where $\ma{D}$,
$\ma{P}$, and $\ma{Q}$ are integer matrices and $\ma{D}$ is diagonal.
Both $\ma{P}$ and $\ma{Q}$ are unimodular matrices (their determinants
are $\pm 1$), so they are invertible.  This is similar to a singular
value decomposition for integer matrices.
\begin{knitrout}
\definecolor{shadecolor}{rgb}{0.969, 0.969, 0.969}\color{fgcolor}\begin{kframe}
\begin{alltt}
\hlstd{R> }\hlstd{M} \hlkwb{<-} \hlkwd{matrix}\hlstd{(}\hlkwd{c}\hlstd{(}
\hlstd{+ }   \hlnum{2}\hlstd{,}  \hlnum{4}\hlstd{,}   \hlnum{4}\hlstd{,}
\hlstd{+ }  \hlopt{-}\hlnum{6}\hlstd{,}  \hlnum{6}\hlstd{,}  \hlnum{12}\hlstd{,}
\hlstd{+ }  \hlnum{10}\hlstd{,} \hlopt{-}\hlnum{4}\hlstd{,} \hlopt{-}\hlnum{16}
\hlstd{+ }\hlstd{),} \hlkwc{nrow} \hlstd{=} \hlnum{3}\hlstd{,} \hlkwc{byrow} \hlstd{=} \hlnum{TRUE}\hlstd{)}
\hlstd{R> }
\hlstd{R> }\hlstd{mats} \hlkwb{<-} \hlkwd{snf}\hlstd{(M)}
\hlstd{R> }\hlstd{P} \hlkwb{<-} \hlstd{mats}\hlopt{$}\hlstd{P; D} \hlkwb{<-} \hlstd{mats}\hlopt{$}\hlstd{D; Q} \hlkwb{<-} \hlstd{mats}\hlopt{$}\hlstd{Q}
\hlstd{R> }
\hlstd{R> }\hlstd{P} \hlopt{%*%} \hlstd{M} \hlopt{%*%} \hlstd{Q}                \hlcom{# = D}
\end{alltt}
\begin{verbatim}
     [,1] [,2] [,3]
[1,]   12    0    0
[2,]    0    6    0
[3,]    0    0    2
\end{verbatim}
\begin{alltt}
\hlstd{R> }\hlkwd{solve}\hlstd{(P)} \hlopt{%*%} \hlstd{D} \hlopt{%*%} \hlkwd{solve}\hlstd{(Q)}  \hlcom{# = M}
\end{alltt}
\begin{verbatim}
     [,1] [,2] [,3]
[1,]    2    4    4
[2,]   -6    6   12
[3,]   10   -4  -16
\end{verbatim}
\begin{alltt}
\hlstd{R> }\hlkwd{det}\hlstd{(P)}
\end{alltt}
\begin{verbatim}
[1] 1
\end{verbatim}
\begin{alltt}
\hlstd{R> }\hlkwd{det}\hlstd{(Q)}
\end{alltt}
\begin{verbatim}
[1] -1
\end{verbatim}
\end{kframe}
\end{knitrout}

\section{Applications}\label{sec:applications}

To say linear algebra is used in many applications is a vast
understatement -- it is the basic mathematics that drives virtually
every real-world application. It provides solutions to problems that
arise both naturally as linear problems as well as linear
approximations to nonlinear problems, e.g.\ Taylor approximations.
Moreover, numerical linear algebra is a very mature technology.
Nonlinear algebra also has many applications, some of which are found
in naturally appearing nonlinear algebraic problems and others as
better-than-linear approximations to non-algebraic nonlinear problems.
However, symbolic and numerical computational solutions are far less
developed for nonlinear algebra than for linear algebra. 

In this section we illustrate how \pkg{m2r} can be used to address two
nonlinear algebraic problems prototypical of statistical problems
amenable to algebraic investigation. Both examples exclusively use
symbolic techniques from commutative algebra/algebraic geometry. We do
not include any examples from the field of numerical algebraic
geometry because, while those methods are both exceedingly powerful
and accessible with \pkg{m2r} via its connections to software such as
\pl{PHCpack} and \pl{Bertini}, they (1) work in fundamentally
different ways than the methods described in Section~\ref{sec:theory}
and (2) are not native to \pl{Macaulay2}. The following examples are
intentionally simple to demonstrate the usefulness of \pkg{m2r} in
addressing nonlinear algebraic problems while not getting bogged down
by a more complex setting.



\subsection{Solving nonlinear systems of algebraic equations}

\begin{figure}[h!]
\begin{center}
\includegraphics[scale=.31]{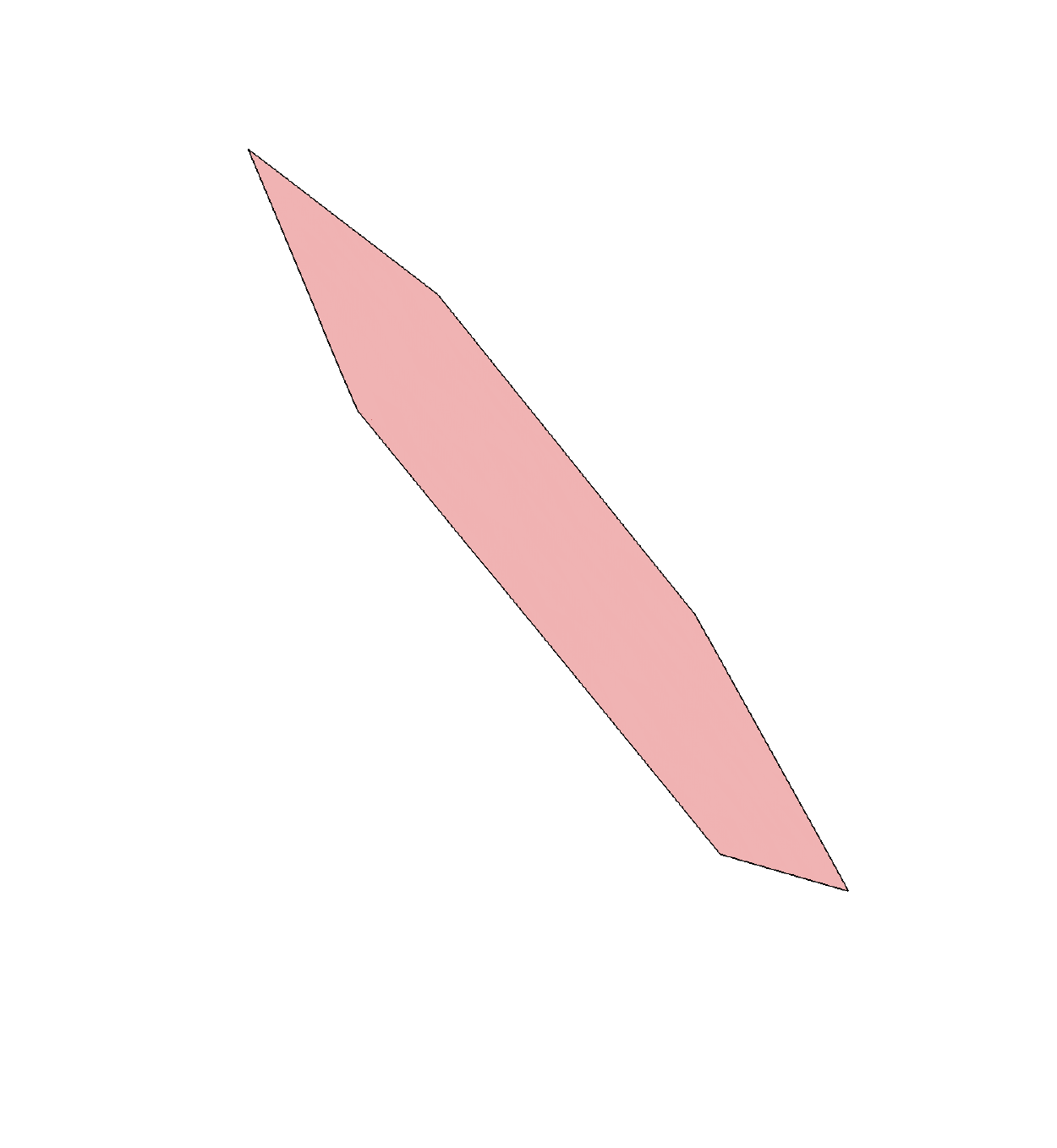}
\includegraphics[scale=.31]{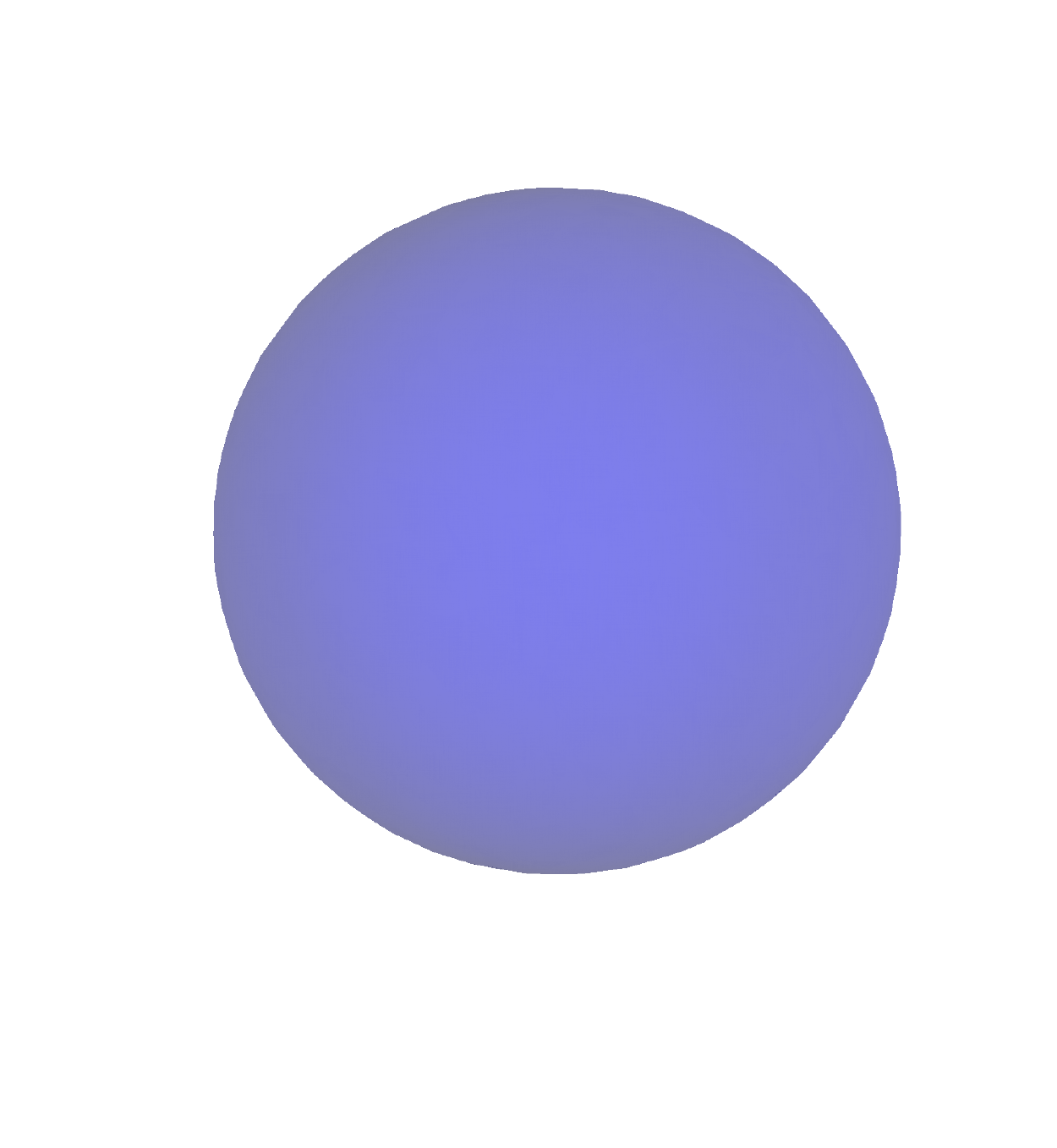}
\includegraphics[scale=.31]{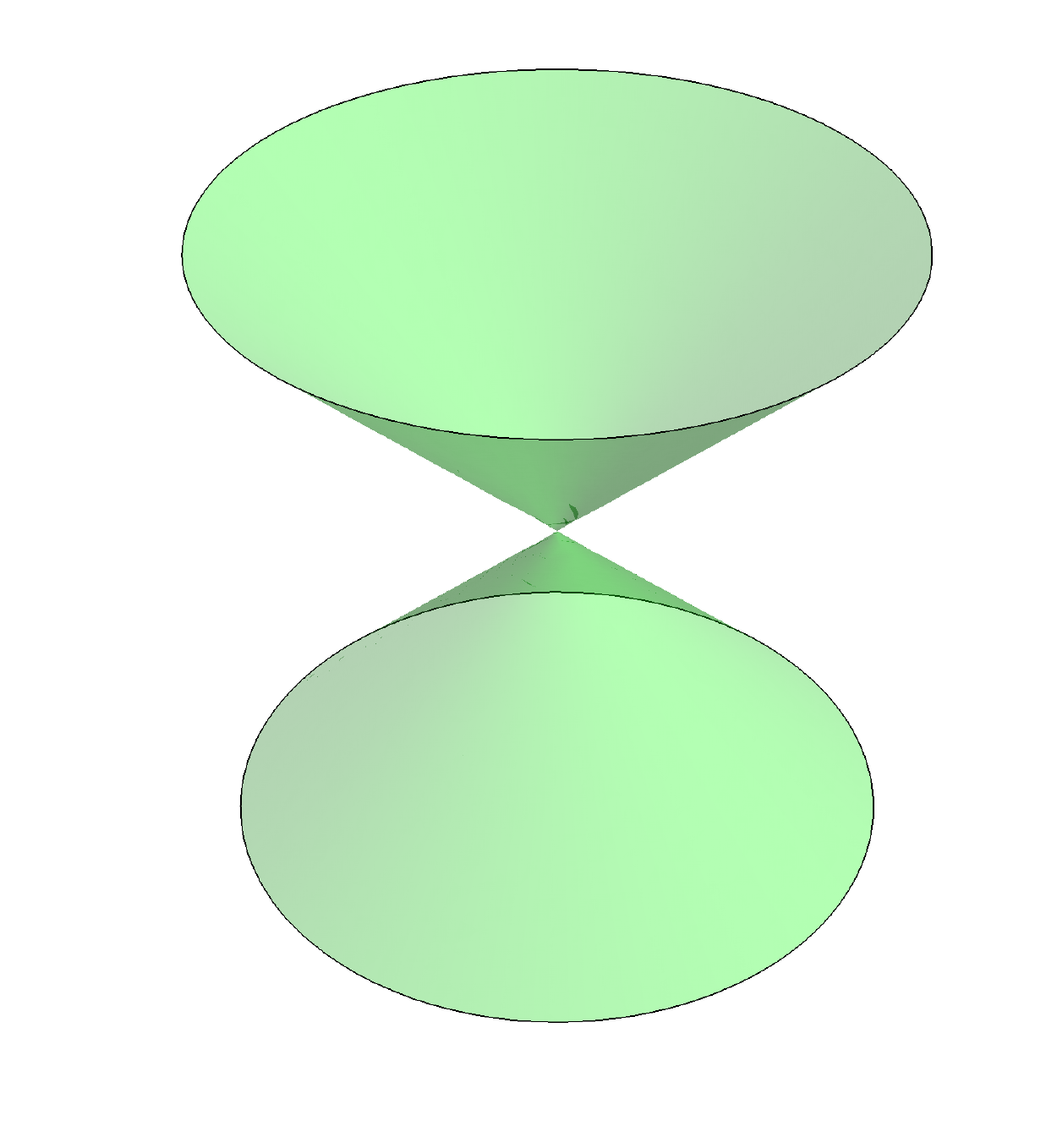}
\includegraphics[scale=.31]{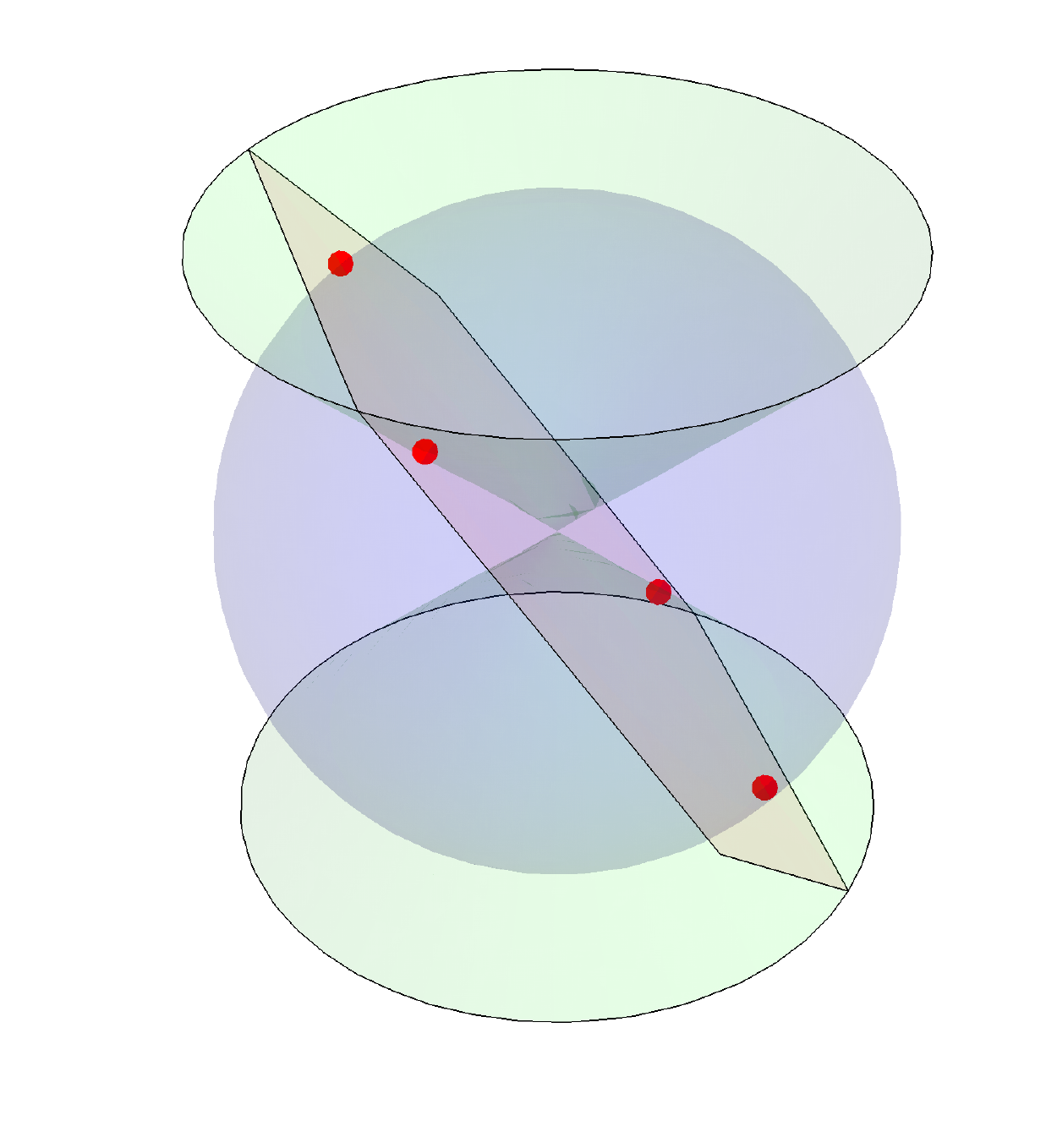}
\end{center}
\caption{The varieties, over $\R$, corresponding to (\ref{eq:varty1}),
	(\ref{eq:varty2}), and (\ref{eq:varty3}) (respectively), and
	their intersection. Solution sets of nonlinear algebraic
	systems consisting of a finite number of points can be
	computed using Gr\"{o}bner bases by recursively finding the
	roots of univariate polynomials.}\label{fig:vartys}
\end{figure}

In this example we show how Gr\"{o}bner bases can be used to solve
zero-dimensional systems of polynomial equations.  Consider the system 
\begin{eqnarray}
x + y + z    &=& 0     \label{eq:varty1} \\
x^{2} + y^{2} + z^{2} &=& 9      \label{eq:varty2} \\
x^{2} + y^{2}         &=& z^{2}  \label{eq:varty3} 
\end{eqnarray}
\noindent Over $\R$, geometrically the variety $\varty{x+y+z, x^{2} +
y^{2} + z^{2} - 9, x^{2} + y^{2} - z^{2}}$, the solution set of $(x,
y, z)$ triples that satisfy (\ref{eq:varty1})--(\ref{eq:varty3}),
corresponds to the intersection of the solution sets of triples that
satisfy each of them individually, i.e. their individual varieties.
These are displayed in Figure~\ref{fig:vartys}. 
 
\pkg{m2r} can be used to find all the solutions to this system exactly
using Gr\"{o}bner bases:
\begin{knitrout}
\definecolor{shadecolor}{rgb}{0.969, 0.969, 0.969}\color{fgcolor}\begin{kframe}
\begin{alltt}
\hlstd{R> }\hlkwd{ring}\hlstd{(}\hlstr{"x"} \hlstd{,}\hlstr{"y"}\hlstd{,} \hlstr{"z"}\hlstd{,} \hlkwc{coefring} \hlstd{=} \hlstr{"QQ"}\hlstd{)}
\end{alltt}
\begin{verbatim}
M2 Ring: QQ[x,y,z], grevlex order
\end{verbatim}
\begin{alltt}
\hlstd{R> }\hlstd{I} \hlkwb{<-} \hlkwd{ideal}\hlstd{(}\hlstr{"x + y + z"}\hlstd{,} \hlstr{"x^2 + y^2 + z^2 - 9"}\hlstd{,} \hlstr{"x^2 + y^2 - z^2"}\hlstd{)}
\hlstd{R> }\hlstd{(grobner_basis} \hlkwb{<-} \hlkwd{gb}\hlstd{(I))}
\end{alltt}
\begin{verbatim}
x  +  y  +  z
2 z^2  -  9
y^2  +  y z
\end{verbatim}
\end{kframe}
\end{knitrout}
\noindent Notice that this system has one polynomial that only
involves $z$, one that only involves $z$ and $y$, and one that
involves $z$, $y$, and $x$.  This is an example of the kind of
nonlinear generalization of Gaussian elimination referred to in
Section~\ref{sec:theory}.

Once \pl{Macaulay2} computes a Gr\"{o}bner basis, it is fairly
straightforward to script a basic solver for nonlinear algebraic
systems that recursively solves the univariate problems and plugs the
solutions into the other equations to obtain other univariate
problems.  In general, when a problem can be reduced to determining
the roots of a univariate polynomial, it is considered solved
\citep{s02}.  An implementation of a univariate polynomial root
finder, the Jenkins-Traub method, is already available in
{\tt \hlstd{base}\hlopt{::}\hlkwd{polyroot}\hlstd{()}}, which \pkg{mpoly} thinly wraps with
{\tt \hlkwd{solve\_unipoly}\hlstd{()}}:
\begin{knitrout}
\definecolor{shadecolor}{rgb}{0.969, 0.969, 0.969}\color{fgcolor}\begin{kframe}
\begin{alltt}
\hlstd{R> }\hlcom{# create simple helpers}
\hlstd{R> }\hlstd{extract_unipoly} \hlkwb{<-} \hlkwa{function}\hlstd{(}\hlkwc{mpolyList}\hlstd{)} \hlkwd{Filter}\hlstd{(is.unipoly, mpolyList)[[}\hlnum{1}\hlstd{]]}
\hlstd{R> }\hlstd{which_unipoly} \hlkwb{<-} \hlkwa{function}\hlstd{(}\hlkwc{mpolyList}\hlstd{)} \hlkwd{sapply}\hlstd{(mpolyList, is.unipoly)} \hlopt{%>%} \hlstd{which}
\hlstd{R> }
\hlstd{R> }\hlcom{# create solver}
\hlstd{R> }\hlstd{solve_gb} \hlkwb{<-} \hlkwa{function}\hlstd{(}\hlkwc{gb}\hlstd{) \{}

\hlstd{+ }  \hlcom{# extract and solve univariate polynomial}
\hlstd{+ }  \hlstd{poly} \hlkwb{<-} \hlkwd{extract_unipoly}\hlstd{(gb)}
\hlstd{+ }  \hlstd{elim_var} \hlkwb{<-} \hlkwd{vars}\hlstd{(poly)}
\hlstd{+ }  \hlstd{solns} \hlkwb{<-} \hlkwd{solve_unipoly}\hlstd{(poly,} \hlkwc{real_only} \hlstd{=} \hlnum{TRUE}\hlstd{)}

\hlstd{+ }  \hlcom{# if univariate polynomial, return}
\hlstd{+ }  \hlkwa{if}\hlstd{(}\hlkwd{length}\hlstd{(gb)} \hlopt{==} \hlnum{1}\hlstd{)} \hlkwd{return}\hlstd{(}\hlkwd{structure}\hlstd{(}\hlkwd{t}\hlstd{(}\hlkwd{t}\hlstd{(solns)),} \hlkwc{.Dimnames} \hlstd{=} \hlkwd{list}\hlstd{(}\hlkwa{NULL}\hlstd{, elim_var)))}

\hlstd{+ }  \hlcom{# remove unipoly from gb; plug solns into remaining system to make new one}
\hlstd{+ }  \hlstd{gb} \hlkwb{<-} \hlkwd{structure}\hlstd{(gb[}\hlopt{-}\hlkwd{which_unipoly}\hlstd{(gb)[}\hlnum{1}\hlstd{],} \hlkwc{drop} \hlstd{=} \hlnum{FALSE}\hlstd{],} \hlkwc{class} \hlstd{=} \hlstr{"mpolyList"}\hlstd{)}
\hlstd{+ }  \hlstd{new_systems} \hlkwb{<-} \hlkwd{lapply}\hlstd{(solns,} \hlkwa{function}\hlstd{(}\hlkwc{soln}\hlstd{)} \hlkwd{plug}\hlstd{(gb, elim_var, soln))}

\hlstd{+ }  \hlcom{# solve reduced system}
\hlstd{+ }  \hlstd{low_solns_list} \hlkwb{<-} \hlkwd{lapply}\hlstd{(new_systems, solve_gb)}
\hlstd{+ }  \hlstd{lower_var_names} \hlkwb{<-} \hlkwd{colnames}\hlstd{(low_solns_list[[}\hlnum{1}\hlstd{]])}

\hlstd{+ }  \hlcom{# aggregate solutions and return}
\hlstd{+ }  \hlkwd{Map}\hlstd{(cbind, solns, low_solns_list)} \hlopt{%>%} \hlkwd{do.call}\hlstd{(}\hlstr{"rbind"}\hlstd{, .)} \hlopt{%>%}
\hlstd{+ }    \hlkwd{structure}\hlstd{(}\hlkwc{.Dimnames} \hlstd{=} \hlkwd{list}\hlstd{(}\hlkwa{NULL}\hlstd{,} \hlkwd{c}\hlstd{(elim_var, lower_var_names)))}

\hlstd{+ }\hlstd{\}}
\end{alltt}
\end{kframe}
\end{knitrout}
\noindent The solver can then be applied to the system
{\tt \hlstd{grobner\_basis}} returned by {\tt \hlkwd{gb}\hlstd{()}} to compute the
solutions to (\ref{eq:varty1})--(\ref{eq:varty3}), the points of
intersection of their corresponding varieties.  We note that the
solver above looks at the variety over $\R$, which is a field
extension of $\Q$, the coefficient ring of the polynomial ring used.
\begin{knitrout}
\definecolor{shadecolor}{rgb}{0.969, 0.969, 0.969}\color{fgcolor}\begin{kframe}
\begin{alltt}
\hlstd{R> }\hlstd{(solns} \hlkwb{<-} \hlkwd{solve_gb}\hlstd{(grobner_basis))} \hlopt{%>%}
\hlstd{+ }  \hlkwd{structure}\hlstd{(}\hlkwc{.Dimnames} \hlstd{=} \hlkwd{list}\hlstd{(}\hlkwd{paste}\hlstd{(}\hlstr{"Soln"}\hlstd{,} \hlnum{1}\hlopt{:}\hlnum{4}\hlstd{,} \hlstr{":"}\hlstd{),} \hlkwd{c}\hlstd{(}\hlstr{"z"}\hlstd{,}\hlstr{"y"}\hlstd{,}\hlstr{"x"}\hlstd{)))}
\end{alltt}
\begin{verbatim}
                z        y        x
Soln 1 :  2.12132  0.00000 -2.12132
Soln 2 :  2.12132 -2.12132  0.00000
Soln 3 : -2.12132  0.00000  2.12132
Soln 4 : -2.12132  2.12132  0.00000
\end{verbatim}
\end{kframe}
\end{knitrout}

In closed form, the four solutions for $(x,y,z)$ are
$\pm\f{3}{\sqrt{2}}(1, 0, -1)$ and $\pm\f{3}{\sqrt{2}}(0, 1, -1)$.
Note that $\f{3}{\sqrt{2}} \approx 2.12132$.  These solutions can be
easily checked by evaluating the original list of polynomials
(\ref{eq:varty1}), (\ref{eq:varty2}), and (\ref{eq:varty3}).
Moreover, the solutions printed above are accurate to 14 digits:
\begin{knitrout}
\definecolor{shadecolor}{rgb}{0.969, 0.969, 0.969}\color{fgcolor}\begin{kframe}
\begin{alltt}
\hlstd{R> }\hlstd{f} \hlkwb{<-} \hlkwd{as.function}\hlstd{(grobner_basis,} \hlkwc{varorder} \hlstd{=} \hlkwd{c}\hlstd{(}\hlstr{"z"}\hlstd{,}\hlstr{"y"}\hlstd{,}\hlstr{"x"}\hlstd{),} \hlkwc{vector} \hlstd{=} \hlnum{TRUE}\hlstd{)}
\hlstd{R> }\hlkwd{apply}\hlstd{(solns,} \hlnum{1}\hlstd{, f)} \hlopt{%>%} \hlkwd{apply}\hlstd{(}\hlnum{2}\hlstd{, round,} \hlkwc{digits} \hlstd{=} \hlnum{14}\hlstd{)} \hlopt{%>%}
\hlstd{+ }  \hlkwd{structure}\hlstd{(}\hlkwc{.Dimnames} \hlstd{=} \hlkwd{list}\hlstd{(}\hlkwd{paste}\hlstd{(}\hlstr{"Eqn"}\hlstd{,} \hlnum{5}\hlopt{:}\hlnum{7}\hlstd{,} \hlstr{":"}\hlstd{),} \hlkwd{paste}\hlstd{(}\hlstr{"Soln"}\hlstd{,} \hlnum{1}\hlopt{:}\hlnum{4}\hlstd{)))}
\end{alltt}
\begin{verbatim}
        Soln 1 Soln 2 Soln 3 Soln 4
Eqn 5 :      0      0      0      0
Eqn 6 :      0      0      0      0
Eqn 7 :      0      0      0      0
\end{verbatim}
\end{kframe}
\end{knitrout}
\noindent We note that a simple numerical strategy that uses
general-purpose optimization routines to solve the system by
minimizing the sum of the squares of the system not only finds only
one solution but is also only correct to 3 digits:
\begin{knitrout}
\definecolor{shadecolor}{rgb}{0.969, 0.969, 0.969}\color{fgcolor}\begin{kframe}
\begin{alltt}
\hlstd{R> }\hlstd{r} \hlkwb{<-} \hlkwa{function}\hlstd{(}\hlkwc{v}\hlstd{) \{}
\hlstd{+ }  \hlstd{x} \hlkwb{<-} \hlstd{v[}\hlnum{1}\hlstd{]; y} \hlkwb{<-} \hlstd{v[}\hlnum{2}\hlstd{]; z} \hlkwb{<-} \hlstd{v[}\hlnum{3}\hlstd{]}
\hlstd{+ }  \hlstd{(x} \hlopt{+} \hlstd{y} \hlopt{+} \hlstd{z)}\hlopt{^}\hlnum{2} \hlopt{+} \hlstd{(x}\hlopt{^}\hlnum{2} \hlopt{+} \hlstd{y}\hlopt{^}\hlnum{2} \hlopt{+} \hlstd{z}\hlopt{^}\hlnum{2} \hlopt{-} \hlnum{9}\hlstd{)}\hlopt{^}\hlnum{2} \hlopt{+} \hlstd{(x}\hlopt{^}\hlnum{2} \hlopt{+} \hlstd{y}\hlopt{^}\hlnum{2} \hlopt{-} \hlstd{z}\hlopt{^}\hlnum{2}\hlstd{)}\hlopt{^}\hlnum{2}
\hlstd{+ }\hlstd{\}}
\hlstd{R> }\hlkwd{optim}\hlstd{(}\hlkwd{c}\hlstd{(}\hlkwc{x} \hlstd{=} \hlnum{0}\hlstd{,} \hlkwc{y} \hlstd{=} \hlnum{0}\hlstd{,} \hlkwc{z} \hlstd{=} \hlnum{0}\hlstd{), r)}\hlopt{$}\hlstd{par}
\end{alltt}
\begin{verbatim}
            x             y             z 
-2.1212603679  0.0002124893  2.1212744693 
\end{verbatim}
\end{kframe}
\end{knitrout}
\noindent This problem is typically dramatically worse in real-world
scenarios with more polynomials of higher degrees.

Though simple, in principle this application can be generalized to any
system of nonlinear algebraic equations.  With appropriate saturation,
it can be generalized even further to systems of rational equations,
i.e. systems involving ratios of multivariate polynomials.  Saturation
is key here because the basic strategy of clearing denominators, i.e.
multiplying equations through by the least common multiple of the
denominators to convert them into polynomial equations, typically
introduces solutions where the original system was previously
undefined.  For example, the system $(\f{y}{x} = 1, y = x^2)$ can be
cleared to $(y = x, y = x^2)$, which suggests the solutions $(0, 0)$
and $(1, 1)$; but $(0,0)$ cannot be a solution since the original
system's first equation ($\f{y}{x} = 1$) is not satisfied at $(0,0)$.
Saturation removes this kind of problem.

New solvers are always of value to the \pl{R} ecosystem, especially
paradigmatically new solvers such as this Gr\"{o}bner basis solution.
One can imagine applications in disparate areas of statistics:
computing estimators via estimating equations (including method of
moments, maximum likelihood, and others), solving polynomial and
rational optimization problems using Lagrange multipliers, and more.
That being said, the Gr\"{o}bner bases method has very definite
limitations: the best algorithms are known to have worst-case behavior
that is doubly-exponential in the number of variables, and solving
systems of polynomial equations is in general known to be an NP-hard
problem.  



\subsection{Independence and nonlinear algebra}

One of the focal application domains of algebraic tools in statistics
is the analysis of multiway contingency tables
\citep{drton2009lectures, aoki2012markov}.  This is for several
reasons.  First, discrete probability distributions, often represented
with probability mass functions in statistics, can be represented as
algebraic objects: non-negative vectors that sum to one.  The ``sum to
one'' condition is a polynomial constraint on the vector of
probabilities.  Second, the definition of independence is an algebraic
condition, as we will see below.  Third, commutative algebra,
particularly combinatorial commutative algebra, has many connections
to integer lattices and polyhedral geometry, which is discussed a
little more at the very end of this example.

A simple example of the algebraic structure of independence is
provided by a two-way contingency table with variables $X$ and $Y$ and
joint distribution $\PRRV{X = x, Y = y} =: p_{xy}$.  If $X$ and $Y$
are both binary so that the sample space of both is $\mc{S}_{X} =
\mc{S}_{Y} = \set{0,1}$, the situation is a $2 \times 2$ table, and
the probabilities are typically denoted $p_{00}$, $p_{01}$, $p_{10}$,
and $p_{11}$.  Collectively, these can be written in order as the
column vector $\ve{p} \in \R^{4}$ that must satisfy the condition
\begin{equation}
\ve{1}_{4}'\ve{p} \ \ = \ \ p_{00} + p_{01} + p_{10} + p_{11} \ \ = \ \ 1.
\end{equation}
If $X$ and $Y$ are independent, the joint distribution factors as a
product of the marginals
\begin{equation}
p_{xy} 
\ \ = \ \ \PRRV{X = x, Y = y} 
\ \ = \ \ \Big(\sum_{y'}\PRRV{X = x, Y = y'}\Big)\Big(\sum_{x'}\PRRV{X = x', Y = y}\Big) 
\ \ =: \ \ p_{x+}p_{+y}.
\end{equation}
Explicitly, independence demands four polynomial constraints of the
probabilities:
\begin{eqnarray}
p_{00} &=& (p_{00} + p_{01}) (p_{00} + p_{10}) \label{eq:indep1} \\
p_{01} &=& (p_{00} + p_{01}) (p_{01} + p_{11}) \label{eq:indep2} \\
p_{10} &=& (p_{10} + p_{11}) (p_{00} + p_{10}) \label{eq:indep3} \\
p_{11} &=& (p_{10} + p_{11}) (p_{01} + p_{11}). \label{eq:indep4} 
\end{eqnarray}

These conditions, along with the sum condition, are routinely
summarized by statisticians in various ways: the log odds-ratio is
zero ($\log \f{p_{00}/p_{01}}{p_{10}/p_{11}} = 0$), the odds-ratio is
one ($\f{p_{00}/p_{01}}{p_{10}/p_{11}} = 1$), or the cross-product
difference is zero ($p_{00}p_{11} - p_{01}p_{10} = 0$)
\citep{agre:2002}. This last condition can be used to derive the other
two.  The distillation of (\ref{eq:indep1})--(\ref{eq:indep4}) to the
more simple cross-product condition $p_{00}p_{11} - p_{01}p_{10} = 0$
can be systematically obtained through the process of computing a
Gr\"{o}bner basis.  This can be done with {\tt \hlkwd{gb}\hlstd{()}}:

\begin{knitrout}
\definecolor{shadecolor}{rgb}{0.969, 0.969, 0.969}\color{fgcolor}\begin{kframe}
\begin{alltt}
\hlstd{R> }\hlkwd{ring}\hlstd{(}\hlstr{"p00"}\hlstd{,} \hlstr{"p01"}\hlstd{,} \hlstr{"p10"}\hlstd{,} \hlstr{"p11"}\hlstd{,} \hlkwc{coefring} \hlstd{=} \hlstr{"QQ"}\hlstd{)}
\end{alltt}
\begin{verbatim}
M2 Ring: QQ[p00,p01,p10,p11], grevlex order
\end{verbatim}
\begin{alltt}
\hlstd{R> }\hlstd{indep_ideal} \hlkwb{<-} \hlkwd{ideal}\hlstd{(}
\hlstd{+ }  \hlstr{"p00 - (p00 + p01) (p00 + p10)"}\hlstd{,}    \hlcom{# p00 = (p00 + p01) (p00 + p10)}
\hlstd{+ }  \hlstr{"p01 - (p00 + p01) (p01 + p11)"}\hlstd{,}    \hlcom{# p01 = (p00 + p01) (p01 + p11)}
\hlstd{+ }  \hlstr{"p10 - (p10 + p11) (p00 + p10)"}\hlstd{,}    \hlcom{# p10 = (p10 + p11) (p00 + p10)}
\hlstd{+ }  \hlstr{"p11 - (p10 + p11) (p01 + p11)"}\hlstd{,}    \hlcom{# p11 = (p10 + p11) (p01 + p11)}
\hlstd{+ }  \hlstr{"p00 + p01 + p10 + p11 - 1"}         \hlcom{# simplex condition}
\hlstd{+ }\hlstd{)}
\hlstd{R> }\hlkwd{gb}\hlstd{(indep_ideal)}
\end{alltt}
\begin{verbatim}
p00  +  p01  +  p10  +  p11  -  1
p01 p10  +  p01 p11  +  p10 p11  +  p11^2  -  p11
\end{verbatim}
\end{kframe}
\end{knitrout}
\noindent Note that the last equation is the one of interest:
\begin{equation}
p_{01}p_{10} + p_{01}p_{11} + p_{10}p_{11} + p_{11}^{2} - p_{11} 
  \ \ = \ \ p_{01}p_{10} + (p_{01} + p_{10} + p_{11} - 1)p_{11}
  \ \ = \ \ p_{01}p_{10} - p_{00}p_{11}.
\end{equation}

In addition to the specification of the model, \pl{Macaulay2} can use
algebraic techniques to determine the dimension of the variety
corresponding to the ideal:
\begin{knitrout}
\definecolor{shadecolor}{rgb}{0.969, 0.969, 0.969}\color{fgcolor}\begin{kframe}
\begin{alltt}
\hlstd{R> }\hlkwd{dimension}\hlstd{(indep_ideal)}
\end{alltt}
\begin{verbatim}
[1] 2
\end{verbatim}
\end{kframe}
\end{knitrout}
\noindent It is well-known that the asymptotic distribution of many
test statistics (e.g.\ Pearson's $\chi^{2}$, the likelihood-ratio
$G^{2}$, etc.) depends on the difference between the dimension of the
saturated model, which is the dimension of the simplex, and the
dimension of the model.  In this case, that distribution is
$\chi^{2}_{\nu}$, where $\nu$ is the difference. The dimension of the
saturated model is $4-1 = 3$, where one degree of freedom is lost to
the simplex condition.  (This can also be checked with
{\tt \hlkwd{dimension}\hlstd{(}\hlkwd{ideal}\hlstd{(}\hlstr{"p00 + p01 + p10 + p11 - 1"}\hlstd{))}}.)  Thus, the
asymptotic distribution of those test statistics is $\chi^{2}_{3-2} =
\chi^{2}_{1}$, which is consistent with the presentation in
introductory courses.

While this example is restricted to independence in the $2 \times 2$
case, it generalizes fully to not only $r \times c$ tables but also to
the multiway case and conditional independence models, a large class
that subsumes graphical models and hierarchical loglinear models.
Partial independence models, where conditional independence statements
do not hold for every level, are also included in this description, as
are conditional independence models with structural zeros.  In short,
working directly with the enumerated polynomial conditions implied by
independence and conditional independence statements expands the
horizons of discrete multivariate analysis.  This also has
ramifications for computing estimators (see \cite{kahle2011minimum}
for details).

One of the most well-developed areas of the young field of algebraic
statistics is that of Markov bases.  Imprecisely, a Markov basis is a
collection of contingency tables called \defn{moves} that, when added
to a given contingency table, result in another contingency table with
the same marginals.  Marginals can be meant in the ordinary sense of
row and column sums for two-way tables, or in a more generalized sense
for more complex models on multiway tables. Given a Markov basis, in
principle one can easily construct a Markov chain Monte Carlo (MCMC)
algorithm to sample from any distribution on the set of tables with
the same marginals as the given table, a set called the \defn{fiber}
of the table. This in turn can be used to generalize Fisher's exact
test, which is used to test for independence in $2 \times 2$ tables,
to any discrete exponential family model on any multiway table, an
enormous generalization. A foundational result in algebraic statistics
called the Fundamental Theorem of Markov Bases implies that Markov
bases can be computed as Gr\"{o}bner bases of a special ideal
\citep{diaconis-sturmfels}.  While \pkg{latter}'s connection to
\pl{4ti2} allows for these kinds of computations, \pkg{m2r}'s
{\tt \hlkwd{gb}\hlstd{()}} gives the user much more flexibility in these kinds of
computations, albeit at significantly reduced performance
\citep{latter, 4ti2}.

\section{Internals and design philosophy}\label{sec:internals}

The \pkg{m2r} package was designed with three basic principles in
mind: (1) make \pl{Macaulay2} as \pl{R}-user friendly as possible, (2)
be as flexible with \pl{Macaulay2} syntax and data structures
possible, and (3) minimize computational overhead.  We advance these
goals with a functional approach by including new data structures, a
robust \pl{Macaulay2} parser, lazy parsing, and reference functions.
In this section we describe these in just enough detail to explain how
they work at a basic level.  For more information, we direct the
reader to the GitHub page at
\href{https://github.com/coneill-math/m2r}{https://github.com/coneill-math/m2r}.


\subsection{\pkg{m2r} data structures}\label{sec:structures}

One of the challenges of working with a computer algebra system in
\pl{R} is that \pl{R} has no infrastructure to handle algebraic
objects.  \pkg{mpoly} alleviates this, but only for polynomials.
There is still a world of other algebraic objects, such as those
described in Section~\ref{sec:theory}, that are represented in
computer algebra systems but do not have any natural analogue in the
\pl{R} ecosystem.

In Section~\ref{sec:parser} we describe how \pkg{m2r} converts
\pl{Macaulay2} data into \pl{R} objects; however, before that
discussion it helps to have an understanding of what kinds of objects
\pkg{m2r} parses \pl{Macaulay2} code into.  Most objects parsed from
\pl{Macaulay2} back into \pl{R} are S3 objects whose last class type
is \code{"m2"} and whose other class types describe the object in
decreasing order of specificity.  For example:
\begin{knitrout}
\definecolor{shadecolor}{rgb}{0.969, 0.969, 0.969}\color{fgcolor}\begin{kframe}
\begin{alltt}
\hlstd{R> }\hlkwd{str}\hlstd{(R)}
\end{alltt}
\begin{verbatim}
Classes 'm2_polynomialring', 'm2'  atomic [1:1] NA
  ..- attr(*, "m2_name")= chr "m2rintring00000001"
  ..- attr(*, "m2_meta")=List of 3
  .. ..$ vars    :List of 4
  .. .. ..$ : chr "t"
  .. .. ..$ : chr "x"
  .. .. ..$ : chr "y"
  .. .. ..$ : chr "z"
  .. ..$ coefring: chr "QQ"
  .. ..$ order   : chr "grevlex"
\end{verbatim}
\end{kframe}
\end{knitrout}
\noindent Created in Section~\ref{sec:basics}, {\tt \hlstd{R}} represents
the polynomial ring $\Q[t,x,y,z]$.  As algebraic objects, rings have
no natural analogue in \pl{R}, so \pkg{m2r} needs to provide a data
structure to represent them.  {\tt \hlstd{R}} is an S3 object of class
{\tt \hlkwd{c}\hlstd{(}\hlstr{"m2\_polynomialring"}\hlstd{,} \hlstr{"m2"}\hlstd{)}}. The value of the object is
{\tt \hlnum{NA}}, a {\tt \hlkwd{logical}\hlstd{(}\hlnum{1}\hlstd{)}} vector; this prevents \pl{R}
users from naively operating on the ring itself. \pkg{m2r} typically
represents algebraic objects by parsing them into \pl{R} as
{\tt \hlnum{NA}} with two attributes, a name ({\tt \hlstd{m2\_name}}) and a
list of metadata ({\tt \hlstd{m2\_meta}}). Both have accessor functions:
\begin{knitrout}
\definecolor{shadecolor}{rgb}{0.969, 0.969, 0.969}\color{fgcolor}\begin{kframe}
\begin{alltt}
\hlstd{R> }\hlkwd{m2_name}\hlstd{(R)}
\end{alltt}
\begin{verbatim}
[1] "m2rintring00000001"
\end{verbatim}
\begin{alltt}
\hlstd{R> }\hlkwd{m2_meta}\hlstd{(R)} \hlopt{%>%} \hlstd{str}
\end{alltt}
\begin{verbatim}
List of 3
 $ vars    :List of 4
  ..$ : chr "t"
  ..$ : chr "x"
  ..$ : chr "y"
  ..$ : chr "z"
 $ coefring: chr "QQ"
 $ order   : chr "grevlex"
\end{verbatim}
\end{kframe}
\end{knitrout}
\noindent The {\tt \hlstd{m2\_name}} attribute is the \pl{Macaulay2}
variable binding for the object; it's the name of the object on the
\pl{Macaulay2} side.  The {\tt \hlstd{m2\_meta}} attribute contains other
information about the object for easy \pl{R} referencing.

Almost every object returned by \pkg{m2r} functions behaves this way
with one major exception: when the object has a natural analogue in
\pl{R}. For example, both \pl{R} and \pl{Macaulay2} have integers and
integer matrices, so it makes sense that when a \pl{Macaulay2} integer
matrix is parsed back into \pl{R}, \pl{R} users can manipulate it just
like an ordinary \pl{R} integer matrix. And that is in fact what
\pkg{m2r} parses the object into, but \pkg{m2r} makes sure that the
object retains the knowledge that it is a \pl{Macaulay2} object.  For
example, the integer matrix {\tt \hlstd{P}} created in the Smith normal
form example in Section~\ref{sec:other} is such an object:
\begin{knitrout}
\definecolor{shadecolor}{rgb}{0.969, 0.969, 0.969}\color{fgcolor}\begin{kframe}
\begin{alltt}
\hlstd{R> }\hlstd{P}
\end{alltt}
\begin{verbatim}
     [,1] [,2] [,3]
[1,]    1    0    1
[2,]    0    1    0
[3,]    0    0    1
M2 Matrix over ZZ[]
\end{verbatim}
\begin{alltt}
\hlstd{R> }\hlkwd{str}\hlstd{(P)}
\end{alltt}
\begin{verbatim}
 int [1:3, 1:3] 1 0 0 0 1 0 1 0 1
 - attr(*, "class")= chr [1:3] "m2_matrix" "m2" "matrix"
 - attr(*, "m2_name")= chr ""
 - attr(*, "m2_meta")=List of 1
  ..$ ring:Classes 'm2_polynomialring', 'm2'  atomic [1:1] NA
  .. .. ..- attr(*, "m2_name")= chr "ZZ"
  .. .. ..- attr(*, "m2_meta")=List of 3
  .. .. .. ..$ vars    : NULL
  .. .. .. ..$ coefring: chr "ZZ"
  .. .. .. ..$ order   : chr "grevlex"
\end{verbatim}
\end{kframe}
\end{knitrout}
\noindent This is what allowed us to compute its determinant directly
in Section~\ref{sec:other} with {\tt \hlkwd{det}\hlstd{(P)}}.


\subsection{The \pl{Macaulay2} parser}\label{sec:parser}

Each call to \pl{Macaulay2} via the {\tt \hlkwd{m2}\hlstd{()}} function produces a
string representing a \pl{Macaulay2} object.  This string, returned
from the \code{toExternalString} function in \pl{Macaulay2}, consists
of valid \pl{Macaulay2} syntax used to recreate the object it
represents, analogous to \pl{R}'s {\tt \hlkwd{dput}\hlstd{()}}. Though this string
is useful for subsequent \pl{Macaulay2} calls because \pl{Macaulay2}
understands it, it typically needs to be parsed in order to be useful
to the \pl{R} user. This task is tedious to do by hand and requires an
understanding of \pl{Macaulay2} syntax.  

{\tt \hlkwd{m2\_parse}\hlstd{()}} is \pkg{m2r}'s general-purpose parsing function.
It takes as input a string of \pl{Macaulay2} output (such as one
returned from \code{toExternalString}) and returns a corresponding
object in \pl{R}. For example, given a string produced from passing a
\pl{Macaulay2} matrix to \pl{Macaulay2}'s \code{toExternalString},
{\tt \hlkwd{m2\_parse}\hlstd{()}} returns a native \pl{R} matrix as part of the
larger {\tt \hlkwd{c}\hlstd{(}\hlstr{"m2\_matrix"}\hlstd{,} \hlstr{"m2"}\hlstd{)}} data structure.

The parser is one of the primary features of \pkg{m2r}.  It was
designed to be as extensible as possible, so that new features could
be added easily and quickly.  For example, in order to add support for
the \pl{Macaulay2} type \code{ideal}, which is returned from
\pl{Macaulay2} as a string of the form
\begin{center}
\verb!ideal map((R)^1,(R)^{{-3},{-3},{-3}},{{a*b*c-d*e*f, a*c*e-b*d*f}})!,
\end{center}
the user simply implements {\tt \hlkwd{m2\_parse\_function.m2\_ideal}\hlstd{()}}, a
single method for the {\tt \hlkwd{m2\_parse\_function}\hlstd{()}} S3 generic that
the parser calls when it encounters an \code{ideal} object.  This
particular function is built in:
\begin{knitrout}
\definecolor{shadecolor}{rgb}{0.969, 0.969, 0.969}\color{fgcolor}\begin{kframe}
\begin{alltt}
\hlstd{R> }\hlstd{m2_parse_function.m2_ideal} \hlkwb{<-} \hlkwa{function}\hlstd{(}\hlkwc{x}\hlstd{) \{}
\hlstd{+ }  \hlkwd{m2_structure}\hlstd{(}
\hlstd{+ }    \hlkwc{m2_name} \hlstd{=} \hlstr{""}\hlstd{,}
\hlstd{+ }    \hlkwc{m2_class} \hlstd{=} \hlstr{"m2_ideal"}\hlstd{,}
\hlstd{+ }    \hlkwc{m2_meta} \hlstd{=} \hlkwd{list}\hlstd{(}\hlkwc{rmap} \hlstd{= x[[}\hlnum{1}\hlstd{]])}
\hlstd{+ }  \hlstd{)}
\hlstd{+ }\hlstd{\}}
\end{alltt}
\end{kframe}
\end{knitrout}
\noindent {\tt \hlkwd{m2\_structure}\hlstd{()}} accepts five arguments:
{\tt \hlstd{x}}, the value of the returned object that is defaulted to
{\tt \hlnum{NA}}; {\tt \hlstd{m2\_name}}, the name of the object;
{\tt \hlstd{m2\_class}}, the higher precedent class; {\tt \hlstd{m2\_meta}}, the
list of metadata; and {\tt \hlstd{base\_class}}, for higher order classes.
In general specific {\tt \hlkwd{m2\_parse\_function}\hlstd{()}} methods accept a
list of arguments {\tt \hlstd{x}} to the \pl{Macaulay2} function
{\tt \hlkwd{ideal}\hlstd{()}}; in this case this consists of a single one-row
matrix object.  When the method is dispatched as part of
{\tt \hlkwd{m2\_parse}\hlstd{()}}, the parser has already parsed the
\code{map(...)} substring to construct an \pl{R}-matrix whose entries
are \pkg{mpoly} objects and passed this via {\tt \hlstd{x[[}\hlnum{1}\hlstd{]]}}.  The
returned {\tt \hlstd{m2\_structure}} thus encapsulates the
{\tt \hlstd{m2\_ideal}} object and has a list of \pkg{mpoly} objects as its
metadata for each polynomial generator of the ideal.  

The recursive nature of the parser effectively black-boxes most of its
inner workings, so that adding new features does not require a deep
understanding of the parser's internal structure (e.g. the tokenizer).
Indeed, much of the currently supported \pkg{m2r} functionality
(including matrix and ideal objects) uses functions like these, built
directly into the parser.  This high level of extensibility ensures
that adding new features is quick and uniform, while requiring as
little additional code as possible. Its simplicity also encourages
contributions from other developers through the \pkg{m2r} GitHub page.


\subsection{Lazy parsing and reference functions}\label{sec:reference}

As noted in Section~\ref{sec:intro}, one of the primary benefits of
\pl{Macaulay2} is its efficiency with large algebraic computations.
For instance, some Gr\"{o}bner basis computations can take many hours
and produce output consisting of several thousand polynomials or
polynomials with several thousand terms.  The \pl{Macaulay2} user can
specify properties to return or have the output immediately passed
into another function.

In order to avoid the computational overhead of copying and parsing
large data structures into \pl{R}, only to then convert them back to
\pl{Macaulay2} for subsequent function calls, nearly every \pkg{m2r}
function has two versions: a reference version and a value version.
Until now, every \pkg{m2r} function we have seen has been the value
version.  As a general naming convention, the reference version of a
function is the value version's name followed by a dot.  For example,
{\tt \hlkwd{gb.}\hlstd{()}} is the reference function corresponding to the value
function {\tt \hlkwd{gb}\hlstd{()}}.

So what is the difference?  Unlike value functions, reference
functions return a pointer to a \pl{Macaulay2} data structure, an S3
object of class {\tt \hlkwd{c}\hlstd{(}\hlstr{"m2\_pointer"}\hlstd{,}\hlstr{"m2"}\hlstd{)}}.  In general, pointers
are not very helpful on the \pl{R} side; they are difficult to
interpret and have somewhat complex printing methods.  For example,
the reference version of {\tt \hlkwd{gb}\hlstd{()}} has the following output:
\begin{knitrout}
\definecolor{shadecolor}{rgb}{0.969, 0.969, 0.969}\color{fgcolor}\begin{kframe}
\begin{alltt}
\hlstd{R> }\hlkwd{gb.}\hlstd{(I)}
\end{alltt}
\begin{verbatim}
M2 Pointer Object
  ExternalString : map((m2rintring00000004)^1,(m2rintring00000004)^{{-1...
         M2 Name : m2rintgb00000006
        M2 Class : Matrix (Type)
\end{verbatim}
\end{kframe}
\end{knitrout}
\noindent Obviously, the output does not appear particularly useful;
it gives no clues as to what the Gr\"{o}bner basis actually is.
Pointers are used as \pl{R}-side handles for \pl{Macaulay2}-side
objects.

Most of the time, the pointer returned from a reference function is
passed into {\tt \hlkwd{m2\_parse}\hlstd{()}} to produce the corresponding \pl{R}
types. (The exception to this is {\tt \hlkwd{m2}\hlstd{()}} itself, which simply
returns the external string part of the pointer returned by
{\tt \hlkwd{m2.}\hlstd{()}}.) In fact, this is precisely what value versions of
functions do; they thinly wrap reference versions with an
{\tt \hlkwd{m2\_parse}\hlstd{()}} call, occasionally with additional parsing. But
users can also pass pointers directly into nearly any \pkg{m2r}
function and obtain the same output without requiring a
computationally expensive call to {\tt \hlkwd{m2\_parse}\hlstd{()}}.





With this design, a novice user can avoid any confusion associated
with pointers by simply omitting the trailing ``{\tt \hlstd{.}}'' from any
functions they use, and their code will work as expected.  However,
advanced users have the option to save additional overhead by using
the reference functions (those ending in ``{\tt \hlstd{.}}'') when they
intend to immediately pass the output back into another \pkg{m2r}
function.

\section{The \pkg{m2r} cloud}\label{sec:connecting}

Ultimately, every \pkg{m2r} function that uses \pl{Macaulay2} invokes
{\tt \hlkwd{m2.}\hlstd{()}}.  Every time {\tt \hlkwd{m2.}\hlstd{()}} is called, it checks for
a connection to a live \pl{Macaulay2} instance.  If one is not found,
{\tt \hlkwd{start\_m2}\hlstd{()}} is run to initialize the \pl{Macaulay2} session.
In this section we describe how \pkg{m2r} makes this connection
between \pl{R} and \pl{Macaulay2}.  We begin with the basic mechanism
of connection, sockets, and then turn to how these connections support
a cloud computing framework that migrates computations off-site,
enabling \pl{Macaulay2} through \pkg{m2r} for Windows users, among
other things.


\subsection{The socket connection between \pl{R} and a local \pl{Macaulay2} instance}
\label{sec:sockets}

\pkg{m2r} uses \defn{sockets} as the primary form of communication
between concurrent \pl{R} and \pl{Macaulay2} sessions.  A socket is a
low-level transfer mechanism used for interprocess communication.
Sockets are commonly used to send and receive data over the internet,
but they can also be used to transfer data between processes running
on the same machine.  Sockets on a given machine are identified by
their \defn{port} number.  To initiate a connection, one endpoint (the
\defn{server}) must open a port for incoming connections, to which the
other endpoint (the \defn{client}) can then connect.  Communication
through a socket is anonymous; a process need not know the location of
the other endpoint when it connects to the socket, sends and receives
data through the socket, or closes its connection.

The socket setup has two key advantages.  First and foremost, it
enables a single tethered \pl{Macaulay2} session to persist for the
duration of the active \pl{R} session, so any variables or functions
the user defines in \pl{Macaulay2} remain available for future use.
Second, the resulting implementation can be easily extended to run the
\pl{R} and \pl{Macaulay2} sessions on different machines, we explore
this in the next section.  

\begin{figure}[h!]
\centering
  \begin{subfigure}[nooneline]{0.45\textwidth}
  \includegraphics[width=2.75in]{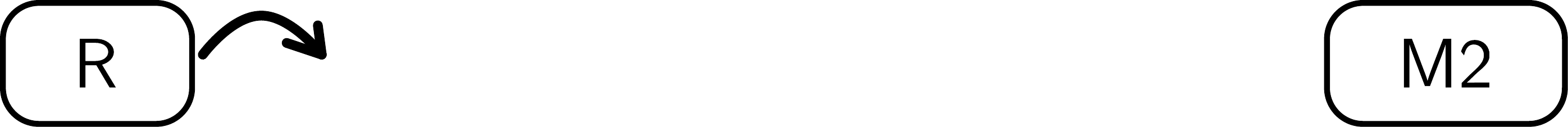}
  \caption{
    \pl{R} begins by launching an \pl{M2} instance, 
    then waits for an available connection on the specified port.
  }\label{fig:socketconnect1}
  \end{subfigure} \hfill
  \begin{subfigure}[nooneline]{0.45\textwidth}
  \includegraphics[width=2.75in]{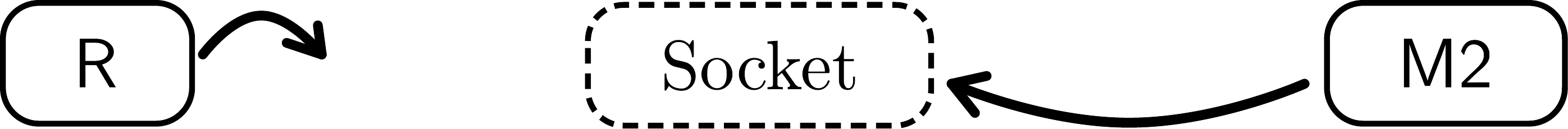}
  \caption{
    Once launched, \pl{M2} opens a socket on the specified port
    and waits for a connection to be established.  
  }\label{fig:socketconnect2}
  \end{subfigure} \\ \vskip .3in
  \begin{subfigure}[nooneline]{0.45\textwidth}
  \includegraphics[width=2.75in]{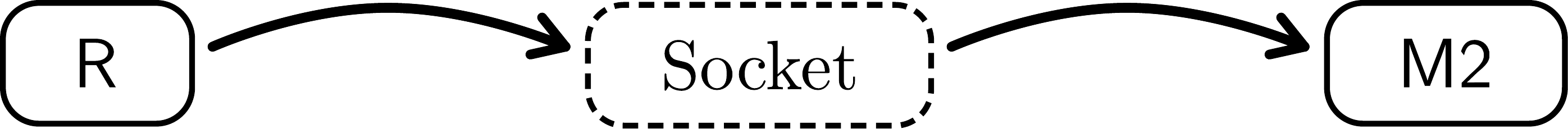}
  \caption{
    \pl{R} connects to the socket, and 
    \pl{M2} pauses while it waits to receive data through the socket.
  }\label{fig:socketconnect3}  
  \end{subfigure} \hfill
  \begin{subfigure}[nooneline]{0.45\textwidth}
  \includegraphics[width=2.75in]{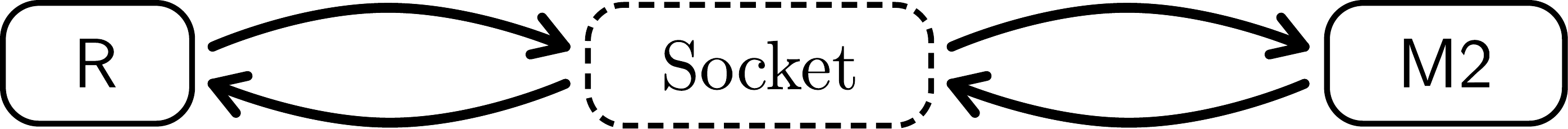}
  \caption{
    Upon successfully connecting to the socket, 
    \pl{R} returns control to the user 
    until {\tt \hlkwd{m2}\hlstd{()}} is called.  
  }\label{fig:socketconnect4}
  \end{subfigure}
\caption{The socket connection process.}
\label{fig:socketconnect}
\end{figure}

\begin{figure}[h!]
\centering
  \begin{subfigure}[p,nooneline]{0.45\textwidth}
  \includegraphics[width=2.75in]{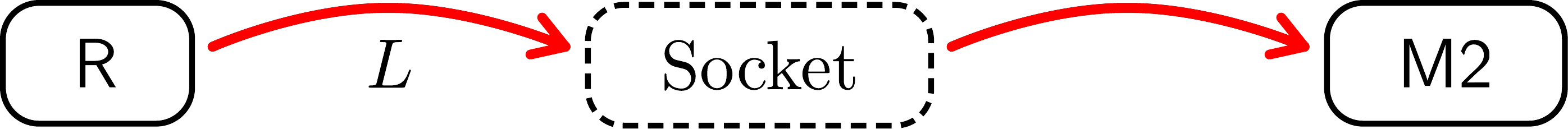} 
  \caption{
    \pl{R} sends a \pl{Macaulay2} source string $L$ through the socket, 
    and then waits for a response from \pl{M2}. 
  }\label{fig:socketsend1} \vfill
  \end{subfigure} \hfill
  \begin{subfigure}[p,nooneline]{0.45\textwidth}
  \includegraphics[width=2.75in]{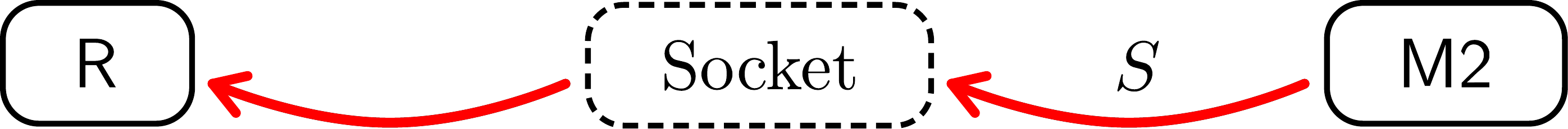}
  \caption{
    \pl{M2} evaluates $L$, sends its response $S$ 
    back through the socket, and then resumes listening.
  }\label{fig:socketsend2}
  \end{subfigure} 
\caption{Messages are passed back and forth through the socket.}
\label{fig:socketsend}
\end{figure}

When {\tt \hlkwd{start\_m2}\hlstd{()}} is called it attempts to initiate a socket
connection between \pl{R} and \pl{Macaulay2} using the sequence of
events documented in Figure~\ref{fig:socketconnect}.  Once \pl{R}
successfully binds to the socket opened by \pl{Macaulay2}, the basic
infrastructure is in place for \pl{R} to send \pl{Macaulay2} code as
character strings to be evaluated; each such code snippet $L$ is
simply relayed to \pl{Macaulay2} through the socket.  After
\pl{Macaulay2} evaluates $L$, it constructs and returns a string $S$
containing (i) any error codes, (ii) the number of lines of output,
and (iii) the output; see Figure~\ref{fig:socketsend} for an
illustration.

After \pl{Macaulay2} issues $S$ it is relayed through the socket to
\pl{R}, which handles any errors and returns the output to the user.
When the \pl{R} session terminates (or {\tt \hlkwd{m2\_stop}\hlstd{()}} is called
by the user), the socket connection is closed by \pl{R} sending an
empty string through the socket signaling end of file (EOF). Upon
receiving an empty string and an EOF signal, \pl{Macaulay2} closes the
socket connection and exits quietly. These steps cleanly kill the
\pl{Macaulay2} process spawned by \pl{R} so that no \pl{Macaulay2}
processes remain orphaned after the \pl{R} session is terminated. 

It is also important to note that the script run by the spawned
\pl{Macaulay2} process does not directly contain any user-supplied
code.  Instead, a \pl{Macaulay2} script that establishes the socket
connection with \pl{R} and conforms to all steps outlined above is
run.


\subsection{\pl{Macaulay2} in the cloud}

Cloud computing as a service has come into prominence in recent years
through the widespread availability of high speed internet connections
and the decreasing cost of hardware and its maintenance at scale,
among other things. In a cloud computing model, the users of a
software system do not need to download the software which they are
using, instead they can simply interact with the software of interest
via a web or terminal interface. Users call on the remote machine to
perform a calculations, and when the remote computations finish the
results are returned to the user. 

The core benefit of a cloud computing model for \pkg{m2r} is that
users no longer have to install \pl{Macaulay2} on their local
machines.  Installing specialized software can be difficult and time
consuming, especially for less computer-savvy users, and this can be
an insurmountable barrier to entry to algebraic statistics and
algebraic methods in general.  This issue is compounded for new users
who are not sure if a certain software is the correct solution for
their problem and so are unwilling to invest the time.  Installing
\pl{Macaulay2} on a Windows machine is an especially arduous task,
creating an enormous barrier to entry for potential Windows users of
the package.  These are common challenges for specialized mathematical
software, and like others before us we concluded that a cloud version
of our software was a worthwhile venture \citep{phcweb2015, habanero}.

Amazon Web Services (AWS, available at \url{https://aws.amazon.com/})
is a subsidiary of Amazon, Inc.\ that sells cloud computing solutions.
AWS's flagship product is the Amazon Elastic Compute Cloud (EC2),
which provides virtual servers of varying performance specs that can
be launched remotely on demand.  To help users get up and running with
\pkg{m2r} and algebraic statistical computing, we have set up a
low-performance EC2 instance dedicated to \pkg{m2r}.  We chose to use
the introductory tier of this product because it suffices for
introducing \pl{R} users to \pl{Macaulay2} and Amazon offers it at no
cost.  It also provides a proof-of-concept model that can be
replicated for a user's own personal cloud.  Instructions for setting
up such an instance can be found on \pkg{m2r}'s GitHub page
(\href{https://github.com/coneill-math/m2r/tree/master/inst/server}{https://github.com/coneill-math/m2r/},
under \code{inst/}).  

A few noteworthy implementation details for remotely running \pkg{m2r}
are in order.  Each remote instance of \pl{Macaulay2} is run within a
virtual machine managed by Docker
(\href{https://www.docker.com}{https://www.docker.com}), an open
source software package that allows for sandboxing of applications
inside distinct lightweight virtual software containers.  Docker
containers provide an additional layer of virtualization that isolates
key resources of the host machine.  This safeguards the host machine
in the sense that nothing executed in a container can affect the host
machine.  Additionally, containers are optimized to be spun up quickly
through efficient usage of host machine resources, significantly
decreasing the time necessary to start a new session and allowing
\pkg{m2r} to connect to on-demand instances of \pl{Macaulay2} in
seconds.

\begin{figure}[h!]
\centering
  \begin{subfigure}[p,nooneline]{0.45\textwidth}
  \includegraphics[width=2.75in]{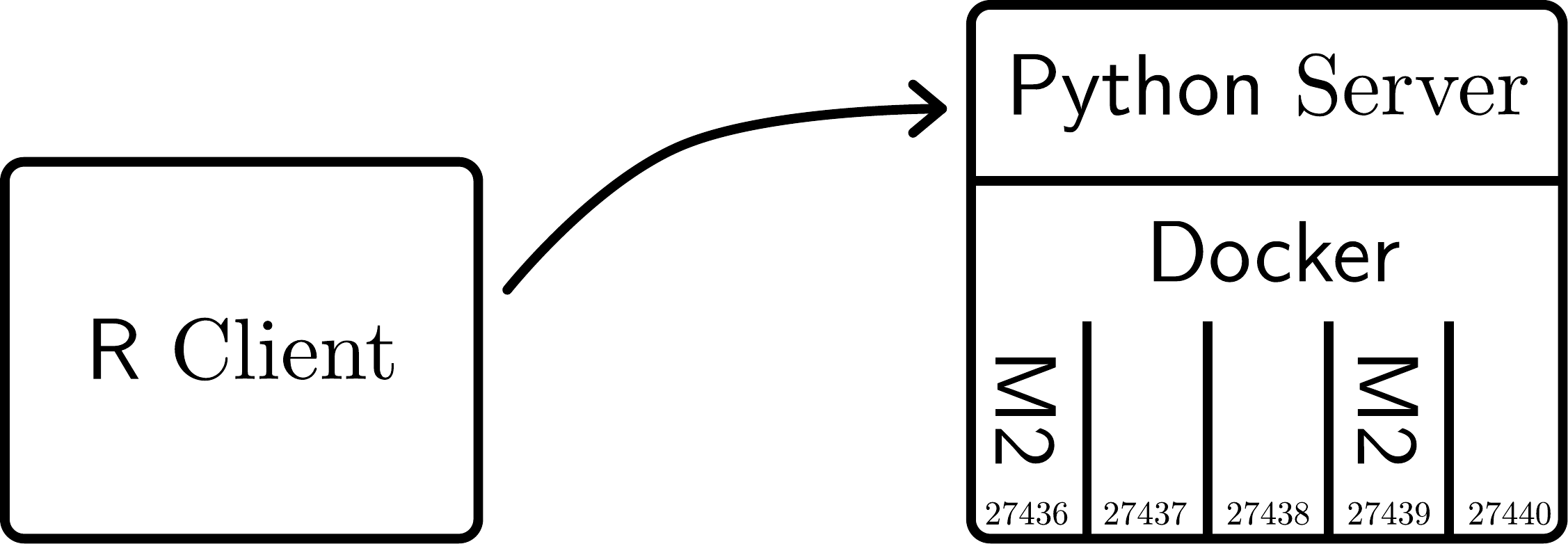} 
  \caption{
    The client \pl{R} session connects to the \pl{Python} server running 
    on the EC2 instance using a static port.  The \pl{Python} server immediately 
    locates an open and unoccupied port $p$ on the EC2 instance. \\ \  
  }\label{fig:socketserver1} \vfill
  \end{subfigure} \hfill
  \begin{subfigure}[p,nooneline]{0.45\textwidth}
  \includegraphics[width=2.75in]{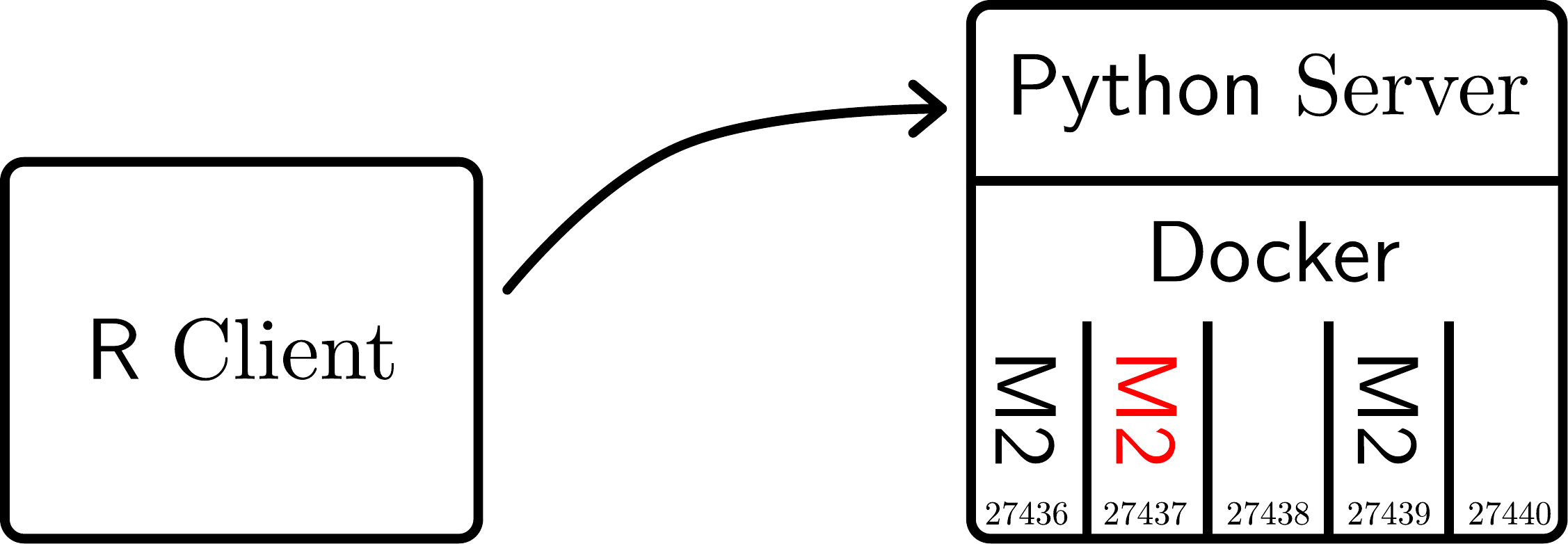}
  \caption{
    The \pl{Python} server launches a new Docker container provisioned with \pl{Macaulay2} and other helpful software. 
    Within this sandboxed container, \pl{Macaulay2} is launched
    and given $p$ as the port number on which to expect an incoming connection.  
  }\label{fig:socketserver2}
  \end{subfigure} \\ \vskip .2in
  \begin{subfigure}[nooneline]{0.45\textwidth}
  \includegraphics[width=2.75in]{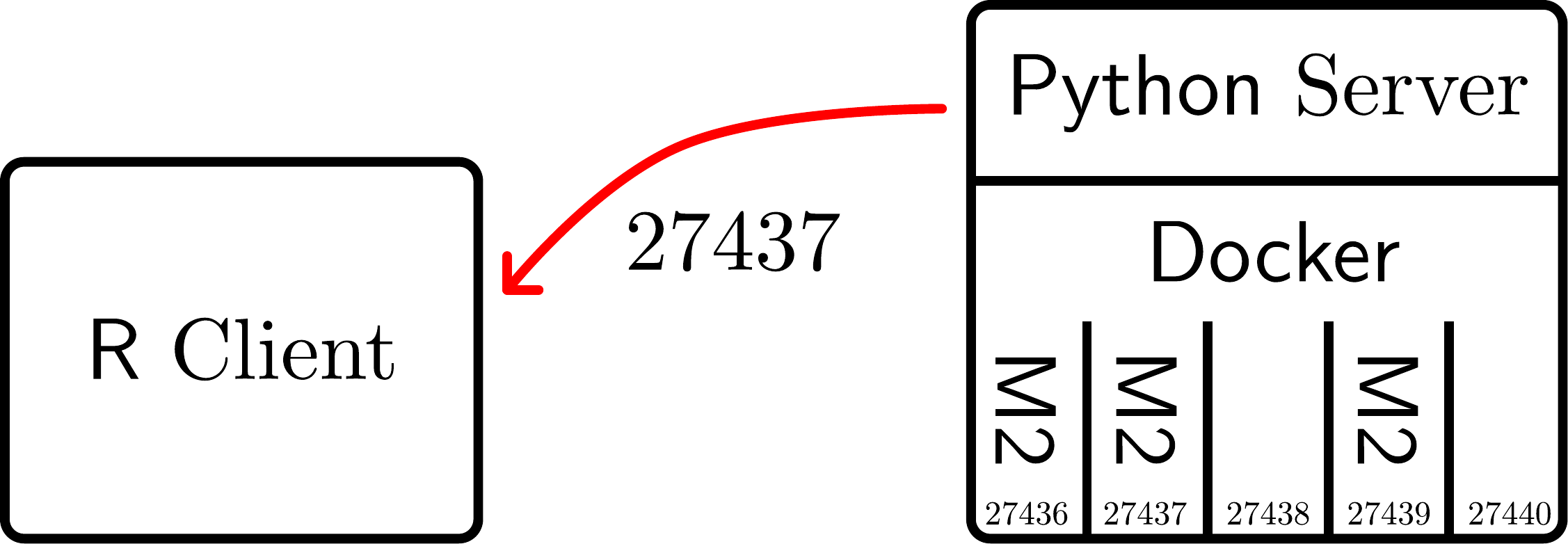}
  \caption{
    The \pl{Python} server sends $p$ to the \pl{R} client, 
    terminates its connection, 
    and begins listening for connections from the next new \pl{R} client.   
  }\label{fig:socketserver3}
  \end{subfigure} \hfill
  \begin{subfigure}[nooneline]{0.45\textwidth}
  \includegraphics[width=2.75in]{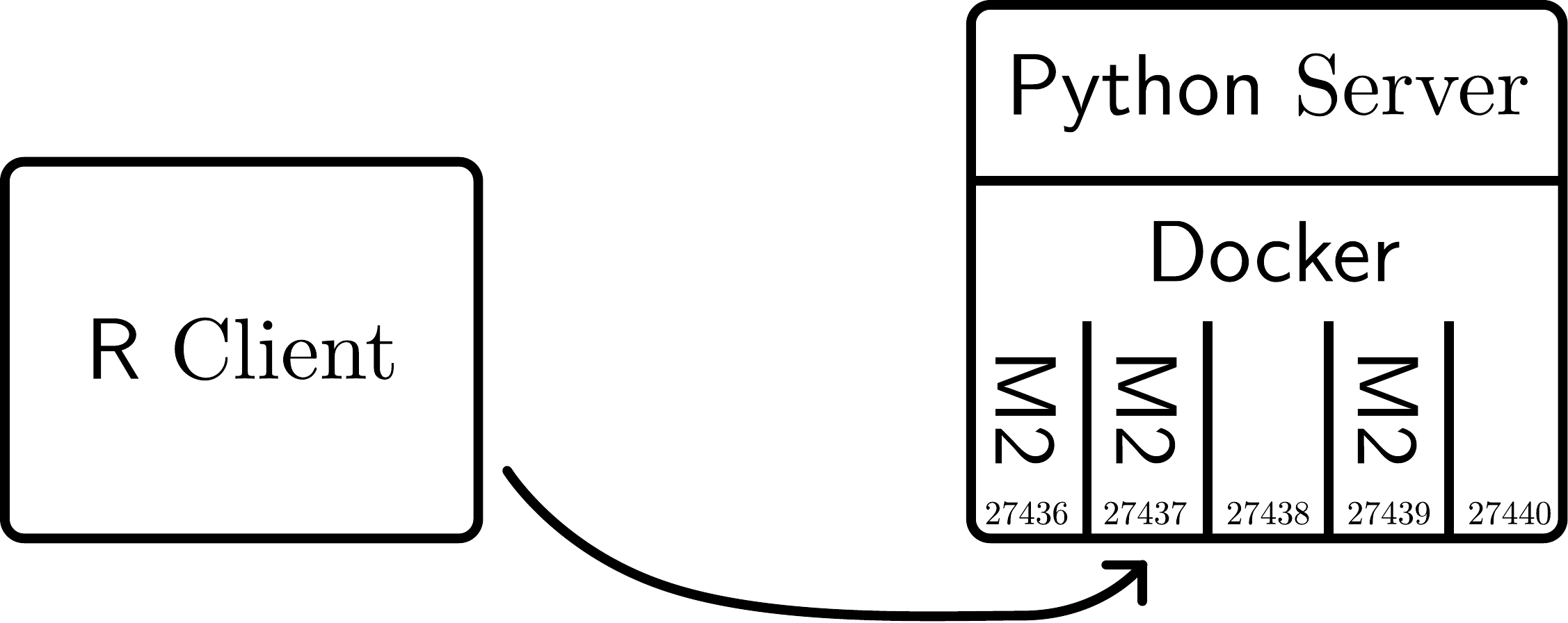}
  \caption{
    The \pl{R} client, upon receiving $p$ from the \pl{Python} server, 
    connects via port $p$ to the \pl{Macaulay2} instance 
    running in the new Docker container
    using the same paradigm used for local 
    \pl{Macaulay2} instances. 
  }\label{fig:socketserver4}
  \end{subfigure} \hfill
\caption{How \pl{R} connects to a \pl{Macaulay2} session on a remote EC2 host.}\label{fig:socketserver}
\end{figure}

While there are many similarities in how \pkg{m2r} connects \pl{R} to
local and remote \pl{Macaulay2} instances, there are some important
differences as well.  Instead of the typical \pkg{m2r} flow where an
instance of \pl{Macaulay2} is launched on the user's local machine,
the server version allows a user to create on-demand \pl{Macaulay2}
instances on an active EC2 instance.  In addition to running and
managing all active Docker containers, the EC2 instance has a
\pl{Python} server script that is used to spawn new Docker instances
and dispatch ports to new clients.  The connection process for a new
\pl{R} client is diagrammed step-by-step in
Figure~\ref{fig:socketserver}. 

The first time {\tt \hlkwd{m2.}\hlstd{()}} is run, \pkg{m2r} will automatically
connect to the cloud if no local \pl{Macaulay2} installation is
detected.  Note that this will always be the case on a Windows
machine, since running a local instance of \pl{Macaulay2} is not
supported.  To bypass a local installation and connect to the cloud,
use the {\tt \hlstd{cloud}} parameter to {\tt \hlkwd{start\_m2}\hlstd{()}}.  
\begin{knitrout}
\definecolor{shadecolor}{rgb}{0.969, 0.969, 0.969}\color{fgcolor}\begin{kframe}
\begin{alltt}
\hlstd{R> }\hlkwd{stop_m2}\hlstd{()}
\hlstd{R> }\hlkwd{start_m2}\hlstd{(}\hlkwc{cloud} \hlstd{=} \hlnum{TRUE}\hlstd{)}
\end{alltt}

{\ttfamily\noindent\itshape\color{messagecolor}{Connecting to M2 in the cloud... }}

{\ttfamily\noindent\itshape\color{messagecolor}{done.}}\begin{alltt}
\hlstd{R> }\hlkwd{m2}\hlstd{(}\hlstr{"1+1"}\hlstd{)}
\end{alltt}
\begin{verbatim}
[1] "2"
\end{verbatim}
\end{kframe}
\end{knitrout}

If the user has the \pl{Macaulay2} server script running on their own
EC2 instance (or any other cloud service for that matter), the URL can
be specified with the {\tt \hlstd{hostname}} parameter to
{\tt \hlkwd{start\_m2}\hlstd{()}}.  From there, everything will work just as if the
user were running a local \pl{Macaulay2} instance.



\section{Future directions}\label{sec:discussion}

In this article we have introduced the new \pkg{m2r} \pl{R} package,
demonstrated several ways it can be used, and explained how it works.
There are several directions of future development that we are excited
about, including performance enhancements for the parser, support for
features such as arbitrary precision numbers and arithmetic with
\pkg{gmp} \citep{gmp, gmpR}, modifications to \pkg{mpoly} for broader
support for multivariate polynomials in \pl{R} (e.g. matrices of
multivariate polynomials), and more. \pl{Macaulay2} boasts a number of
packages for algebraic statistics that are ripe for implementation and
of interest to \pl{R} users and the statistics community more broadly.
We invite collaborators to contact us directly and share their ideas
on the GitHub page.


\section*{Acknowledgements}
The authors would like to thank Robert Harrison for editorial work on
the article.  This material is based upon work supported by the
National Science Foundation under Grant Nos.
\href{https://nsf.gov/awardsearch/showAward?AWD_ID=1321794}{1321794}
and
\href{https://nsf.gov/awardsearch/showAward?AWD_ID=1622449}{1622449}.


\bibliographystyle{elsarticle-harv}
\bibliography{__0-bibliography.bib}

\end{document}